\documentclass{aa}  

\usepackage{natbib}
\usepackage{graphicx}
\usepackage{txfonts}
\usepackage{gensymb}
\usepackage{mathtools}
\usepackage{booktabs}
\usepackage{array}
\usepackage{multirow}
\usepackage{makecell} 
\usepackage{tabularx} 
\newcolumntype{C}[1]{>{\centering\arraybackslash}m{#1}}
\usepackage{xspace}
\newcommand{\BD}{BD+43\degree3654\xspace}
\newcommand{\BN}{BD+60\degree2522\xspace}
\newcommand{\ergs}{erg\,s$^{-1}$\xspace}
\newcommand{\kms}{km\,s$^{-1}$\xspace}
\newcommand{\mJybeam}{\,mJy\,beam$^{-1}$\xspace} 
\newcommand{\Msunyr}{\,M$_\odot$\,yr$^{-1}$\xspace} 

\begin{document}

   \title{Probing the non-thermal physics of stellar bow shocks\\ using radio observations}

   \author{J.~R. Martinez\inst{1,2}
          \and 
          S. del Palacio\inst{3}
          \and
          V. Bosch-Ramon\inst{4}
          }

   \institute{Facultad de Ciencias Exactas, UNLP,
              Calle 47 y 115, CP(1900), La Plata, Buenos Aires, Argentina.\\
              \email{jmartinez@iar.unlp.edu.ar}
         \and
             Instituto Argentino de Radioastronom\'ia (CCT La Plata, CONICET), C.C.5, (1894) Villa Elisa, Buenos Aires, Argentina.
         \and 
            Department of Space, Earth and Environment, Chalmers University of Technology, SE-412 96 Gothenburg, Sweden.
         \and 
             Departament de F\'{i}sica Qu\`antica i Astrof\'{i}sica, Institut de Ci\`encies del Cosmos (ICC), Universitat de Barcelona (IEEC-UB), Mart\'{i} i Franqu\`es 1, E08028 Barcelona, Spain.
             }

   \date{}

 
  \abstract
   {Massive runaway stars produce bow shocks in the interstellar medium. Recent observations revealed radio emission from a few of these objects, but the origin of this radiation remains poorly understood.}
   {We aim to interpret this radio emission and assess under which conditions it could be either thermal (free--free) or non-thermal (synchrotron), and how to use the observational data to infer physical properties of the bow shocks.}
   {We used an extended non-thermal emission model for stellar bow shocks for which we incorporated a consistent calculation of the thermal emission from the forward shock. We fitted this model to the available radio data (spectral and intensity maps), including largely unexplored data at low frequencies. In addition, we used a simplified one-zone model to estimate the gamma-ray emission from particles escaping the bow shocks.}
   {We can only explain the radio data from the best sampled systems (\BD and \BN) assuming a hard electron energy distribution below $\sim$1~GeV, a high efficiency of conversion of (shocked) wind kinetic power into relativistic electrons ($\sim 1-5\%$), and a relatively high magnetic-to-thermal pressure ratio of $\eta_\mathrm{B}\sim 0.2$.
   In the other systems, the interpretation of the observed flux density is more ambiguous, although a non-thermal scenario is also favoured.
   We also show how complementary observations at other frequencies can allow us to place stronger constraints in the model. We also estimated the gamma-ray fluxes from the HII regions around the bow shocks of \BD and \BN, and obtained luminosities at GeV energies of $\sim 10^{33}$~erg\,s$^{-1}$ and $10^{32}$~erg\,s$^{-1}$, respectively, under reasonable assumptions.}
   {Stellar bow shocks can potentially be very efficient particle accelerators. This work provides multi-wavelength predictions of their emission and demonstrates the key role of low-frequency radio observations in unveiling particle acceleration processes. The prospects of detections with next-generation observatories such as SKA and ngVLA are very promising. Finally, \BD may be detected in GeV in the near future, while bow shocks in general may turn out to be non-negligible sources of (at least leptonic) low-energy cosmic rays.}

   \keywords{Radiation mechanisms: non-thermal -- Radiation mechanisms: thermal -- Acceleration of particles -- Shock waves -- Radio continuum: general
               }

   \maketitle
%

\section{Introduction}
    
Stellar winds play an important role in the evolution of massive stars and their feedback in the interstellar medium (ISM). An interesting case of study is when a massive OB star moves with a supersonic velocity with respect to its surrounding medium \citep[see e.g.][for a recent review]{Mackey2022}. In this case, the interaction between the highly supersonic stellar wind and the ISM leads to the formation of a bow shock (BS), which includes a forward shock (FS) that propagates through the ISM and a reverse shock (RS) that propagates through the stellar wind \citep{Weaver_1977}.  

The FS compresses and heats the ISM gas and dust. The heated dust can emit in the infrared, which has been one of the major tracers of stellar BSs \citep{vB_1988, Peri_2012, Peri_2015, Kobulnicky2016}. Moreover, since radiative losses in the FS are usually efficient, the shocked gas can also produce significant free--free emission at radio wavelengths \citep{VdE_2022}, together with optical lines such as H$_\alpha$ \citep{Gull1979}. On the other hand, the RS is fast and adiabatic. Relativistic particles can be accelerated in this shock via diffusive mechanisms. These particles, in turn, can interact with matter and electromagnetic fields emitting broadband non-thermal (NT) radiation. In particular, NT electrons can emit synchrotron emission in the radio band, which was first detected in the BS associated with the star \BD \citep{Benaglia_2010, Benaglia_2021}. The smoking gun of synchrotron radiation is a flux density with a negative spectral index, that is, $S_\nu \propto \nu^\alpha$ with $\alpha<0$. More recently, \cite{Moutzouri_2022} found evidence of NT emission in \BN as well. A few putative radio counterparts were additionally suggested by \cite{Peri_2015} and were further supported by \cite{VdE_2022} using observations from the Rapid ASKAP Continuum Survey (RACS) at 887.5~MHz. MeerKAT observations also revealed a BS in Vela X-1 \citep{VdE2022_VelaX1}. Nonetheless, the nature (thermal or NT) of the radiation in these objects could not be determined directly as it was not possible to derive spectral indices from the observations. The recent observational progress is encouraging, and further breakthroughs are expected with the next generation of radio telescopes such as the Square Kilometre Array (SKA\footnote{\url{https://www.skao.int}.}) and the next-generation Very Large Array (ngVLA\footnote{\url{https://ngvla.nrao.edu}.}) 

Despite the observational progress, we still lack theoretical tools to properly interpret and extract physical information from radio continuum observations. So far only one-zone models have been used to interpret the data from 
the BSs recently detected in radio   \citep{VdE2022_VelaX1, VdE_2022, Moutzouri_2022}, while more sophisticated multi-zone models focused on the NT radiation \citep{del_Valle_2018, del_Palacio_2018, Martinez_2022} have neglected the free--free emission in the radio band. For this reason, in this work we extend the multi-zone model from \cite{del_Palacio_2018} to include this emission component. We apply this generalised model to characterise all the stellar BSs reported in the literature as radio emitters. 

The paper is organised as follows: in Sect.~\ref{sec:systems} we present the sample of studied systems, in Sect.~\ref{sec:model} we describe our thermodynamic and emission model, and in Sects.~\ref{sec:Results} and \ref{sec:conclusions} we present and summarise our main results and conclusions, respectively. 
    

\section{Systems studied}\label{sec:systems}

We focused on all the stellar BSs that have been confirmed as radio sources.
Knowing the velocity of the stars and the distance to the systems is crucial to characterise the BSs. The latest catalogue from the {\it Gaia} mission (Data Release 3) provides the most precise measurements to calculate these parameters, that is, proper motions and parallaxes \citep{Gaia3}. In general, the velocity with respect to the surrounding medium can be separated into two components: one tangential ($V_{\star {\rm t}}$) and one radial ($V_{\star {\rm r}}$). To calculate $V_{\star {\rm t}}$, we transformed the measured stellar proper motions to Galactic coordinates and subtracted the local Galactic velocity field using the Oort constants given by \cite{B_B_2019}. Unfortunately, the radial velocities have large uncertainties. For example, the radial velocities provided by {\it Gaia} are not calibrated for massive stars, and in general the stellar winds produce broad emission lines \citep{C&E_1977}, making it difficult to measure Doppler shifts from the radial motion of the star. As a consequence, the stellar velocities used in this paper are estimations based mostly on the tangential components and the morphology of the BSs.

In Table~\ref{table:systems} we list the systems studied and their most relevant parameters. We focus mainly on \BD and \BN, as these two are confirmed NT sources that have been observed at several radio frequencies. In addition, we explored the more poorly characterised sample of radio BSs reported by \cite{VdE2022_VelaX1} and \cite{VdE_2022} to assess the nature of their emission. Below we provide a short summary of each source.

\subsection{\BD}\label{subsec:BD43}

\BD is a massive 04If star with a supersonic wind of $v_{\rm w} \approx 3\,000~$\kms \citep{Muijres_2012} and $\dot{M}_{\rm w} \approx 4\times 10^{-6}$~\Msunyr \citep{Moutzouri_2022}. The star was ejected from the Cygnus OB2 association \citep{Comeron_2007} and its parallax is $\Pi = 0.58 \pm 0.01~$mas \citep{Gaia3}, yielding a distance of $d = 1.72 \pm 0.02~$kpc. The proper motion parameters are $\left(\mu_\alpha\,\cos{\delta}, \mu_\delta \right) = \left(-2.59\pm 0.01, +0.73\pm 0.01\right)~$mas\,yr$^{-1}$, yielding a tangential velocity of $V_{\star {\rm t}} = 43.8\pm 0.9~$\kms. From the morphology of the BS, we infer that this component of the velocity is predominant (as for a dominant radial component the BS would appear more circular), and we adopted the value suggested by \cite{Benaglia_2021} of $V_\star = 50~$\kms. Taking into account the projected distance from the star to the BS apex $R_{\rm 0, proj} \approx 3.9'$ \citep{Benaglia_2021}, the distance $d$ and the stellar velocity $V_\star$, and assuming a fully ionised medium, we derive an ambient density of $n_{\rm ISM}\approx 6$~cm$^{-3}$ (which includes ionised atoms and free electrons).

The BS of \BD is the first stellar BS to be detected at radio frequencies. \cite{Benaglia_2010} observed this source at 1.4~GHz and 4.86~GHz with the Very Large Array (VLA) and measured an average negative spectral index $\alpha \approx -0.5$. This indicates the presence of synchrotron emission and consequently of NT particles. More recently, \cite{Benaglia_2021} and \cite{Moutzouri_2022} reported new observations of this BS. \cite{Benaglia_2021} presented a deeper image of the BS using the upgraded VLA in the S band (2--4~GHz) and measured fluxes up to a factor of $\sim 10$ larger than the ones from \cite{Benaglia_2010}. 
Furthermore, \cite{Benaglia_2021} reported a steeper spectrum, with a slope of $\alpha \sim -1$. Nevertheless, this could be caused by the loss of flux at higher frequencies by the interferometer, which can be problematic for extended sources embedded in regions with diffuse emission. This effect artificially leads to a steeper spectrum \citep[see e.g.][for a discussion on these issues]{Green2022}. In addition, \cite{Moutzouri_2022} observed the BS with the upgraded VLA in the C (4--8 GHz) and X (8--12 GHz) bands and with the single-dish Effelsberg radio telescope in the C band. Using the C band was an attempt to resolve the issue of the missing flux due to extended structures (as the largest angular scales probed by the VLA at higher frequencies are too small compared with the size of \BD). However, this comes at the expense of having a very large beam size given by the single dish. In the feathered data they report, the fluxes are overestimated due to the contamination of flux from a nearby HII region located north of the BS. Luckily, this HII region has a well-characterised thermal spectrum \citep{Benaglia_2021}, so we can subtract its contribution to the feathered data and thus recover the true flux from the BS. Lastly, we also included as additional constraints the information from the intensity maps from the Westerbork Northern Sky Survey (WENSS) at 325~MHz \citep[][see our Appendix~\ref{appendix:radio_data}]{WENSS_1997} and the non-detection from the TIFR GMRT Sky Survey (TGSS) at 150~MHz \citep{Intema_2017}.  


\subsection{\BN}\label{subsec:BD60}
The massive O star \BN launches a powerful stellar wind with a mass-loss rate of $\dot{M} \approx 10^{-6}$~\Msunyr and a terminal velocity of $v_\infty \approx 2000$~\kms \citep{Moutzouri_2022}. This star is within a high density region of $n_\mathrm{ISM} \sim 27$~cm$^{-3}$, where it moves with a supersonic velocity, giving rise to a prominent BS. This whole structure is known as the Bubble Nebula. The optical image from \cite{Moore2002} revealed the presence of extended regions of diffuse gas, together with some higher density ionised regions to the north and to the west of the central star that are actually not part of the BS. 

The VLA observations of this source in the frequency range of 4--12~GHz revealed the radio emission from the BS \citep{Moutzouri_2022}. The spectral energy distribution (SED) has a spectral index of $\alpha \approx -0.8$, which as in the case of the BS of \BD, is steeper than the canonical value of $\alpha = -0.5$ expected to arise in strong non-relativistic shocks \citep{Bell1978}. We note that there is a small region with $\alpha>0.5$ to the west of \BN, coming from the dense ionised clumps reported by \cite{Moore2002}. This region is bright at 4--12~GHz ($\sim40$\mJybeam) and dominated the single-dish observations presented by \cite{Moutzouri_2022}. This makes the reported flux values unreliable and we therefore model only the observed intensity maps. We also note that significant free--free emission from the ionised, dilute gas in the nebula can also affect observations at gigahertz frequencies \citep{Green2022}. In addition, it is likely that the BS is actually more extended, with fainter emission undetected by the VLA observation at 4--12~GHz, similarly as in the BS of \BD \citep{Benaglia_2021}. Given these limitations, model fitting can only be addressed within order-of-magnitude uncertainties.

\subsection{
G1}\label{subsec:
G1}

G1 is an O7V-type star ejected from the star-forming region NGC 6357. \cite{Gvaramadze_2011} reported infrared observations of its BS, while \cite{VdE_2022} reported observations of the system at 887.5~MHz from the RACS. According to \cite{Peri_2015}, the wind velocity and mass-loss rate of the star are $v_{\rm w} \approx 2100~$\kms and $\dot{M} \approx 2 \times 10^{-7}~$\Msunyr, respectively. The parallax of the star is $\Pi = 0.56 \pm 0.02~$mas \citep{Gaia3}, corresponding to a distance $d = 1.79\pm 0.03~$kpc. The proper motions are $\left(\mu_\alpha\,\cos{\delta}, \mu_\delta \right) = \left(1.50\pm 0.02, -2.27\pm 0.01\right)~$mas yr$^{-1}$, from which we derive a tangential velocity of $V_{\star, {\rm t}} = 11.5 \pm 0.8~$\kms. Given that the sound speed in the ISM is $c_{\rm s} \sim 10~$\kms, and considering the morphology of the BS, the radial component of the velocity must be $V_{\star, {\rm r}} \sim V_{\star, {\rm t}}$ in order to form the structure. Therefore, we consider a stellar peculiar velocity of $V_\star = 20~$\kms.

\subsection{G3}\label{subsec:G3}

Catalogued as an O6Vn -- O5V type star \citep{Drilling_1981, Peri_2015}, G3 was also ejected from NGC 6357 \citep{Gvaramadze_2011} and is also located at a distance of $d = 1.79\pm 0.06$~kpc. \cite{VdE_2022} reported the detection of G3 at 887.5~MHz from the RACS, confirming G3 as a radio-stellar BS. According to \cite{Peri_2015}, the supersonic wind has a velocity of $v_{\rm w} = 2000$~\kms and the mass-loss rate is $\dot{M} = 4\times 10^{-7}$~\Msunyr. The proper motion parameters of G3 are $\left(\mu_\alpha\,\cos{\delta}, \mu_\delta \right) = \left(1.77\pm 0.02, -3.26\pm 0.01\right)~$mas\,yr$^{-1}$, leading to a tangential velocity of $V_{\star,{\rm t}}= 19\pm1$~\kms. Since the environment of G3 might be similar to the environment of G1, we assumed the same ambient density as G3, yielding a peculiar velocity of G1 of $V_\star = 22$~\kms.

\subsection{Vela X-1}\label{subsec:VelaX1}

Vela X-1 is a high-mass X-ray binary, consisting of a neutron star and a supergiant. \cite{VdE2022_VelaX1} observed the system with the MeerKAT telescope and detected radio emission from the BS at 865--1712~MHz. The supergiant star has a dense wind with a mass loss rate of $\dot{M} \approx 10^{-6}$~\Msunyr, and a velocity of $v_{\rm w} \approx 700$~\kms \citep{Grinberg_2017}. Located at a distance of $d = 2.02\pm 0.6$~kpc, the proper motion parameters of Vela-X1 are $\left(\mu_\alpha\,\cos{\delta}, \mu_\delta \right) = \left(-4.82\pm 0.01, 9.28\pm 0.02\right)~$mas~yr$^{-1}$. We derive a tangential velocity of $V_{\star{\rm t}} = 59.7\pm2.2$~\kms and, as the morphology of the BS implies $V_{\star{\rm r}} < V_{\star{\rm t}}$, we consider $V_\star = 65$~\kms \citep[consistent with $V_{\star{\rm r}} \approx 30$~\kms, as inferred by][]{Gvaramadze_2018}. The BS is located at a projected distance of $\approx 0.8'$ from the star (corresponding to a linear distance of $\approx 0.47$~pc).

\begin{table*}[h]    
    \renewcommand{\arraystretch}{1.3} 
    \centering
    \caption[]{Parameters of the systems studied (see Sect.~\ref{sec:systems} for a discussion on their possible uncertainties). (a) \cite{Benaglia_2021}; (b) \cite{Gaia3} (stellar velocities are estimations based on tangential components);  (c) \cite{Peri_2012}; (d) \cite{Sota_2011}; (e) \cite{Peri_2015}; (f) \cite{Houk_1978}; (g) \cite{Muijres_2012}; (h) \cite{Moutzouri_2022}; (i) \cite{Grinberg_2017}; (j) \cite{Toala_2020}. The ambient densities are calculated in terms of the projections $R_{\rm 0, proj}$ in the plane of the sky reported by \cite{Peri_2012}, \cite{Peri_2015}, \cite{Benaglia_2021}, \cite{Moutzouri_2022} and \cite{Gvaramadze_2018}. We assume solar abundances.}
    \begin{tabular}{l l c c c c c}
    \hline\hline\noalign{\smallskip}
    \multirow{2}{*}{Parameter}     & \multirow{2}{*}{Symbol} & \multicolumn{5}{ c}{System} \\ \cline{3-7}
              &         & \BD    & BD+60\degree 2522    & 
    G1    & G3    & Vela X-1\\ 
    \midrule
    Peculiar velocity   & $V_\star$ [km\,s$^{-1}$]    & 50$^{(a)(b)}$           & 42$^{(b)}$     & 20$^{(b)}$    & 22$^{(b)}$   & 65$^{(b)}$\\
    Ambient density     & $n_\mathrm{ISM}$ [cm$^{-3}$] & 6     & 27   & 30    & 30   & 2\\
    Spectral type       &             & O4If$^{(c)}$            & O6.5V$^{(d)}$     & O7V$^{(e)}$   & O5V$^{(e)}$   & B0.5Ia$^{(f)}$\\ 
    Wind velocity       & $v_\mathrm{w}$ [km\,s$^{-1}$] & 3000$^{(g)}$         & 2000$^{(h)}$    & 2100$^{(e)}$  & 2000$^{(e)}$     & 700$^{(i)}$\\
    Wind mass-loss rate & $\dot{M}_{\rm w}$ [M$_\sun$\,yr$^{-1}$] & $4\times10^{-6\,(h)}$ & $1.3\times10^{-6\,(j)}$  & $2.2\times10^{-7\,(e)}$ & $4\times10^{-7\,(e)}$ & $1\times10^{-6\,(i)}$\\ 
    Distance            & $d$ [kpc]  & 1.72$^{(b)}$ & 2.99$^{(b)}$ & 2.02$^{(b)}$ & 1.79$^{(b)}$ & 1.79$^{(b)}$\\
    \bottomrule
    \end{tabular}
\label{table:systems}
\end{table*}


%
\section{Model}\label{sec:model}
%

\begin{figure}[t]
    \centering
    \includegraphics[width=1.0\linewidth]{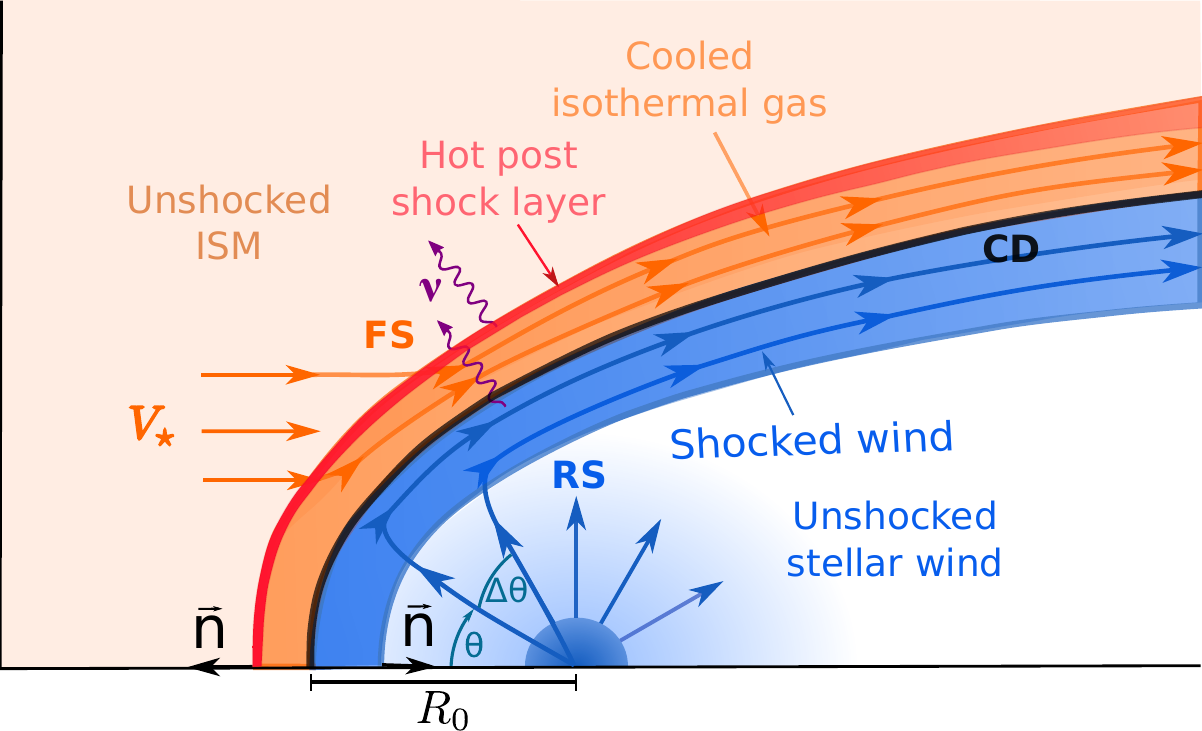}
    \caption{Sketch (not to scale) of a BS of a runaway star. The FS and the RS are separated by the CD, represented by a solid black line. The immediate post-shock external medium, the shocked medium ionised by the star, and the shocked stellar wind are shown in red, orange, and blue, respectively. The solid lines with arrows indicate the shocked gas streamlines, from the injection position up to the tail of the BS. Adapted from \cite{Martinez_2022}.}
    \label{fig:bow_shock}
\end{figure}

In Fig.~\ref{fig:bow_shock} we show a sketch of an archetypal BS of a runaway star. The strong supersonic stellar wind hits the oncoming ISM separating the system into four regions: un-shocked and shocked stellar wind, and un-shocked and shocked ISM; the shocked stellar wind and the shocked ISM are separated by a contact discontinuity (CD). This is a surface defined by the condition that the mass momentum flows tangential to it, so that shocked material from the RS and the FS do not mix (neglecting turbulent effects, assumption that may not be valid far from the BS apex). The model presented here is based on the one developed in \cite{Martinez_2022} and we refer the reader to that article for details. Here we briefly summarise the main aspects of the model and focus on the new upgrades implemented in this work.

The shape of the CD for a star moving through a warm ISM was determined analytically by \cite{Christie_2016}. We approximated each shock as a thin and axisymmetric structure, co-spatial with the CD, and characterised the BS employing a multi-zone model. Shocked gas convects away from the apex at a fixed angle $\phi$ around the symmetry axis. For a given $\phi$, we discretised each shock in multiple cells, corresponding to different angles $\theta$ measured from the star (with $\theta=0$ at the apex of the BS). Finally, we followed the shocked gas and calculated the thermodynamic quantities cell by cell up to a certain angle, $\theta_{\rm max}$, at which the FS vanishes. Although the RS could still accelerate particles further, the shocked flow will tend to develop instabilities, so we assumed that the RS also stops at $\theta_{\rm max}$. The value of $\theta_{\rm max}$ depends mostly on the velocity of the star: the BS extends up to larger values of $\theta_{\rm max}$ for stars with higher velocities.

The closest position of the CD to the star is the stagnation point. It is located along the CD axis of symmetry, and it is the position where the ISM and the stellar wind pressures cancel each other out. The ISM pressure is a combination of ram and thermal pressure, while in the stellar wind, the thermal pressure is negligible compared to its ram pressure. The position of the stagnation point is then given by
\begin{equation}
    R_0 = \sqrt{\frac{\dot{M}v_{\rm w}}{4\pi \Bigl(n_{\rm ISM}k_{\rm B} T_{\rm ISM} + \rho_{\rm ISM}V_\star^2\Bigr)}},
    \label{Eq:R_0}
\end{equation}
where $\rho_{\rm ISM} = n_{\rm ISM}\,\mu_{\rm ISM}\,m_p$ is the density of the ISM, $\mu_{\rm ISM}$ its mean molecular weight, and $m_p$ the proton mass.


\subsection{Reverse shock}

The RS is strong and adiabatic. The shock heats and compresses the stellar wind, also compressing the local magnetic field. Shocked material convects away from the BS apex without cooling, as cooling timescales are much larger than escape timescales. Under such conditions, electrons and protons can be accelerated via diffusive shock acceleration. Electrons emit NT radiation, predominantly synchrotron emission in radio and inverse Compton (IC) emission in X-rays and gamma rays. The IC can be produced by the interaction with the stellar radiation field or the infrared radiation emitted by the heated dust in the FS. On the other hand, proton-proton (p-p) cooling of protons is inefficient, as the shocked wind is very diluted. 

\subsubsection{Hydrodynamics of the reverse shock}\label{subsubsec:hydro_RS}

In the star reference frame, the shocked wind is reaccelerated from the BS apex towards distant regions, moving parallel to the CD. In the regions where the shocked wind is subsonic, we assumed that the total pressure from the upstream medium is transformed into thermal pressure in the downstream medium. In the supersonic regime, instead, we considered that only the total momentum flux component that is perpendicular to the BS is converted to thermal pressure. Moreover, we assumed that the shocked fluid behaves as an adiabatic gas with coefficient $\gamma = 5/3$. Under this hypothesis, we obtain the density using the polytropic relation, $\rho(\theta) \propto P(\theta)^{1/\gamma}$, and specific enthalpy conservation across the shock front.

The magnetic field in the RS can be generated by adiabatic compression of the star magnetic field and/or generated in situ by the diffusion of cosmic rays \citep[e.g.][]{Pittard_2021}. Because of the former, the magnetic field is expected to be parallel to the RS surface \citep{Marcowith2016}. In the subsonic region we calculated it by imposing that the magnetic pressure is a fraction $\eta_{\rm B}$ of the thermal pressure, so
\begin{equation}
    B(\theta) = \left[\eta_{\rm B} 8\pi P(\theta)\right]^{1/2}.
    \label{Eq:B_sub}
\end{equation}
Here $\eta_B < 1$: otherwise, the gas becomes incompressible and diffusive shock acceleration cannot take place. Lastly, in the supersonic region we assumed that the magnetic field remains frozen to the plasma. Then we obtained the magnetic field considering conservation of the magnetic flux:
\begin{equation}
   B(\theta) = B(\theta_\mathrm{c})\, \left(
   \frac{\rho(\theta)}{\rho(\theta_\mathrm{c})}
   \frac{v(\theta_\mathrm{c})}{v(\theta)}
   \right)^{1/2},
   \label{Eq:B_sup}
\end{equation}
with $\theta_{\rm c}$ the angle at which the plasma becomes supersonic.

\subsubsection{Non-thermal particles}\label{subsubsec:NT}

We discretised the RS in multiple cells, corresponding to different angles $\theta$. To compute the NT emission we considered several one-dimensional linear emitters. Each of these starts at a different cell and consists of multiple cells located on the path of the emitter along the shock. We assumed that NT particles are injected in the first cell of each emitter, and we summed up the contributions from all the emitters at a certain angle $\phi$, obtaining a one-dimensional set of emitters \citep[a sketch is shown in Fig.~5 in][]{del_Palacio_2018}. Lastly, we rotated this set around the symmetry axis to get the entire structure of the RS.

We needed to assume an energy distribution of injected NT particles at the {\it i}-th cell. For electrons with a power-law energy distribution with spectral index $p$, their synchrotron spectra is also a power law with spectral index $\alpha = \left(p+1\right)/2$ (with $N\left(E\right) \propto E^p$ and $S \propto \nu^\alpha$). Thus, observations at radio frequencies can be used to derive the injection spectrum. Observations by \cite{Benaglia_2021} and \cite{Moutzouri_2022} revealed emission with $\alpha \sim -(0.75$--$1)$, consistent with an injection index of $\approx p\approx -(2.5$--$3$). However, assuming a constant injection index of $p = -2.5$ (or $p=-3$) for all electron energies leads to unrealistic energetic requirements in the BS of \BN and also to largely over-predicting the flux of the BS of \BD at 150~MHz (see Sect.~\ref{subsec:radio_results} for further details). This suggests that the electron energy distribution has a harder spectral index at lower energies. We therefore propose a piecewise injection spectrum:
\begin{equation}
   Q\left(E,\theta_i \right) = 
   \begin{dcases}       
   Q_0\,E^{p_1} \exp{\left(-E/E_{\rm max}\right)} &     E < E_{\rm break} \sim 2~{\rm GeV} \\
   Q_0\,E^{p_2} \exp{\left(-E/E_{\rm max}\right)} &     E \geq E_{\rm break} \sim 2~{\rm GeV}.
   \end{dcases}
    \label{Eq:Q_inj}
\end{equation}
The value of $p_1$ is poorly constrained by the available observations, so we simply adopted a hard spectrum below $E_{\rm break}$ with $p_1 \sim 1$, since softer values lead to overestimating the emission at 150~MHz. Additionally, we set a soft index $p_2\sim-3$ above $E_{\rm break}$. 

On the other hand, $Q_0$ is a normalisation factor and $E_{\rm max}$ a cut-off energy, both dependent on $\theta$. We set the normalisation factor by the condition
\begin{equation}
    \int_{2m_{\rm e}c^2}^{E_{\rm max}} E\,Q\left(E,\theta_i\right)\,{\rm d}E = \Delta L_{\rm NT}\left(\theta_i\right) = f_{\rm NT}\,\Delta L_\perp\left(\theta_i\right),
    \label{Eq:Q_0}
\end{equation}
where $\Delta L_{\rm NT}\left(\theta_i\right)$ is the power injected into NT particles at the position $\theta_i$. The fraction $f_{\rm NT}$  of the power $\Delta L_\perp\left(\theta_i\right)$ injected perpendicularly into the RS is a free parameter of the model, while $\Delta L_\perp(\theta_i)$ depends on the RS region: 
\begin{equation}
   \Delta L_\perp(\theta_i) =
    \begin{dcases}
    0.5\,\rho_{\rm w}(\theta_i)\,v_{\rm w}^3\,\Delta A_\perp(\theta_i)    \quad {\rm (subsonic\;region)} \\
    0.5\,\rho_{\rm w}(\theta_i)\,v_{\rm w}^2\,v_{{\rm w} \perp}(\theta_i)\,\Delta A_\perp(\theta_i)    \quad {\rm (supersonic\;region)},
    \end{dcases}
    \label{Eq:L_NT}
\end{equation}
with $\Delta A_\perp(\theta_i)$ the area of the {\it i}-th cell perpendicular to ${\bf v_{\rm w}}$. In turn, a fraction $\Delta L_{\rm NT, e}(\theta_i) = f_{{\rm NT,}e}\, \Delta L_\perp(\theta_i)$ is injected into electrons, while the remaining fraction, $f_{{\rm NT,}p} = f_{\rm NT} - f_{{\rm NT,}e}$, is injected into protons. Finally, the cut-off energy is obtained by equating the acceleration and energy timescales.

We notice that the magnetic field plays an important role in the acceleration of particles. Although we supposed that the field lines are mostly parallel to the BS surface, we addressed the acceleration process phenomenologically, and different geometries would lead to similar results. For instance, shock drift acceleration could also take place, but the shape of the energy particle distribution would be unaltered if particles can complete enough acceleration cycles \citep{Matthews2020}.

\begin{figure}[t]
\centering
\includegraphics[angle=270,width=\linewidth]{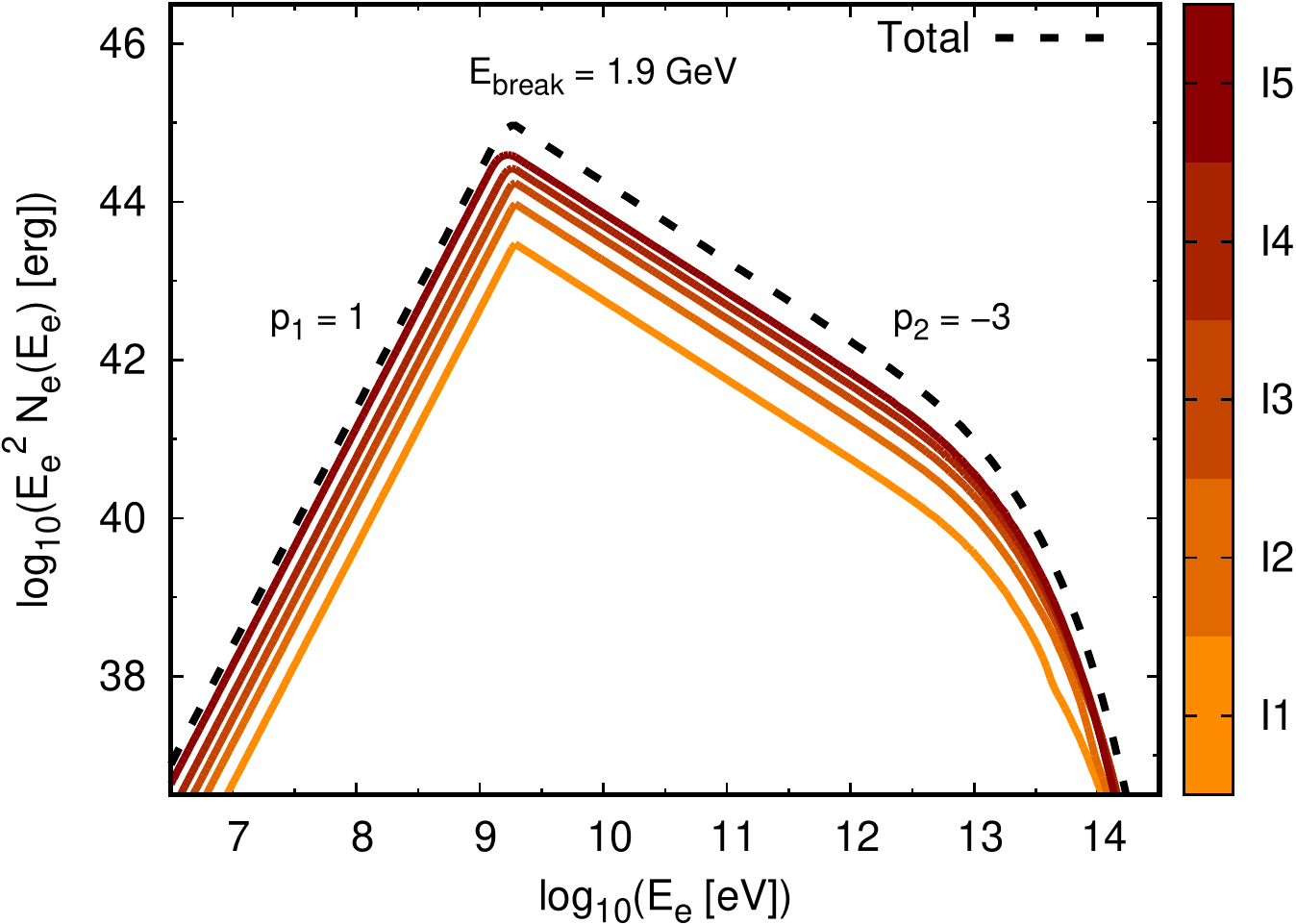}
\caption{Electron energy distribution for the RS of the BS of \BD. We discretise the distribution in five segments of length $\theta_{\rm max}/5$ along the BS (colour scale, with I1 the one starting at the apex of the BS). The dashed black line shows the total distribution. The adopted injection function is given by Eq.~\ref{Eq:Q_inj}.
}
\label{fig:diste_BD_43}
\end{figure}

We approximated the steady-state particle distribution at the injection cell by:
\begin{equation}
    N_0\left(E, \theta_i\right) \approx Q\left(E, \theta_i\right) \times {\rm min}\left(t_{\rm cell}, t_{\rm cool}\right),
    \label{Eq:N_0}
\end{equation}
with $t_{\rm cell}$ the cell convection time and $t_{\rm cool}$ the cooling timescale of NT particles. Lastly, we considered convection by the shocked gas and calculated the particle energy distribution cell by cell assuming conservation of the flux of particles in the space of energy and positions. As an example, in Fig.~\ref{fig:diste_BD_43} we show the resulting electron energy distribution for the BS of \BD.
For further details of the RS treatment, we refer the reader to \cite{Martinez_2022}.

\subsection{Forward shock}\label{sec:FS}

The FS is a radiative shock that compresses and heats the oncoming ISM material. This creates a thin hot post-shock layer where injected power is re-emitted via thermal bremsstrahlung and line emission \citep{Gull1979}. Moreover, early-type runaway stars present wind-driven BSs in which the star photo-ionises the entire BS structure \citep{Henney_2019}. Consequently, there are two layers in the FS: a hot immediate post-shock layer that cools locally, and an isothermal layer (IL) with a temperature $T \sim 8\,000$~K that is kept warm and ionised by the star \citep{Mackey2022}, as shown in Fig.~\ref{fig:bow_shock}. Henceforth we will refer to the former as the hot layer (HL) and the latter as the isothermal layer. We note that, although particles can be accelerated in low-Mach-number shocks, a stellar velocity of hundreds of km~s$^{-1}$ is required for this to be efficient \citep{Martinez_2022}. Thus, we considered only thermal radiation from the FS.


\subsubsection{Hydrodynamics of the forward shock}\label{subsubsec:hydro_FS}

We employed Rankine-Hugoniot jump conditions to determine the thermodynamic quantities in the HL. We calculated the density in terms of the ISM density and the compression factor:
\begin{equation}
    \rho_{\rm HL} (\theta) = \zeta(\theta)\,\rho_{\rm ISM},
    \label{Eq:rho_1}
\end{equation}
and the pressure in terms of the Mach number and the ISM thermal pressure:
\begin{equation}
    P_{\rm HL}(\theta) = \left(\frac{2 \gamma M(\theta)^2-\gamma+1}{\gamma-1}\right)\,P_{\rm th, ISM}.
    \label{Eq:P_1}
\end{equation}
The ISM thermal pressure depends on the ISM temperature and numerical density as $P_{\rm th,ISM} = n_{\rm ISM}k_{\rm B}T_{\rm ISM}$. As runaway stars are ejected from their birth clusters at high velocities and are usually isolated, it is unlikely that other sources could affect the properties of their surrounding ISM. Thus, we focused on the case of wind-driven BSs with Strömgren spheres that are much larger than the BS itself. Under these considerations, we expect an ionised ISM with temperature $T_{\rm ISM} \sim 8\,000~$K  \citep{Weaver_1977}. 

The temperature of the HL is also determined in terms of the Mach number and the ISM temperature:
\begin{equation}
    T_{\rm HL}(\theta) = \frac{\left[\left(\gamma-1\right)M(\theta)^2+2\right]\left[2\gamma M(\theta)^2-\left(\gamma-1\right)\right]}{\left(\gamma+1\right)^2M(\theta)^2}\,T_{\rm ISM}
    \label{Eq:T_1}.
\end{equation}
Using this expression we fixed $\theta_{\rm max}$ as the angle at which $\rho(\theta) \approx \rho_{\rm ISM}$, where the FS vanishes.

Knowing the temperature of the HL enables the calculation of the density of the IL. Given that the pressure at the IL is $P_{\rm IL}(\theta) = P_{\rm HL}(\theta)$ because of pressure equilibrium, the density increases with respect to $\rho_{\rm HL}(\theta)$ by a factor $T_{\rm HL}(\theta)/T_{\rm IL}$:
\begin{equation}
    \rho_{\rm IL}\left(\theta\right) = \frac{P_{\rm HL}(\theta)}{k_{\rm B}T_{\rm IL}}\,\mu\,m_p = \frac{T_{\rm HL}(\theta)}{T_{\rm IL}}\,\rho_{\rm HL}(\theta),
    \label{Eq:rho_2}
\end{equation}
where $\mu\approx  0.6$ is the mean molecular weight (assuming solar abundances).

Lastly, considering mass conservation across the shock, the width of the IL region is:
\begin{equation}
     H_{\rm IL}(\theta) = \eta_{\rm H} \left[ \frac{\int_0^\theta \rho_{\rm ISM}(\theta')\,V_\star\,A_\perp(\theta')\,{\rm d}\theta'}{2\pi\,R(\theta)\,\sin{(\theta)}\,\rho_{\rm IL}(\theta)\,v_\parallel(\theta)} -  \left(\frac{\rho_{\rm HL}(\theta)}{\rho_{\rm IL}(\theta)}\right)H_{\rm HL}(\theta) \right],
     \label{Eq:H_2}
\end{equation}
where $v_\parallel(\theta)$ is the convection velocity of the shocked fluid, and $H_{\rm HL}\left(\theta\right) = \left(V_{\star \perp}/\zeta\right)t_{\rm cool, HL}$ the width of the HL. Finally, $\eta_{\rm H} \leq 1$ is a free parameter that accounts for additional processes not included in the model that can reduce the IL thickness, such as thermal conduction and instabilities \citep[e.g.][]{Comeron_Kaper_1998}.

\subsubsection{Thermal emission}{\label{subsubsec:thermal}}

Thermal radiation from the HL and the IL is optically thin. We calculated their free--free spectral luminosity as \citep[e.g.][]{Rybicki_1986}:
\begin{equation}
    L_{\rm ff}(\theta, \epsilon_{\rm ph}) = 6.38 \times 10^{-38} \frac{L_0\,Z^2\,n_{\rm e}\,n_{\rm i}\,g_{\rm ff}\,\epsilon_{\rm ph}}{\sqrt{T}}\,\exp{\left(-\frac{\epsilon_{\rm ph}}{k_{\rm B}T}\right)}\;{\rm erg\,s^{-1}},
    \label{Eq:L_ff}
\end{equation}
where all quantities are in cgs units. In this formula, $Z$ is the mean atomic number (assumed to be $Z = 0.79$ for solar abundances and typical conditions of an HII region), $g_{\rm ff}$ the mean Gaunt factor, $\epsilon_{\rm ph}$ the energy of the photons, $n_{\rm i} \approx n_{\rm e} \approx n/2$ the ion and electron number densities, respectively, and $L_0$ a normalisation factor. 

The total free--free luminosity from the HL depends mostly on the velocity of the star and the density of the ISM, and it is normalised at each cell by the condition:
\begin{equation}
    L_{\rm ff, HL}(\theta_i) = \frac{\Lambda_{\rm ff}(T)}{\Lambda_{\rm tot}(T)}\left(1-\frac{T_{\rm IL}}{T_{\rm HL}(\theta_i)}\right)\, L_\perp(\theta_i),
    \label{Eq:L_ff_1}
\end{equation}
with $L_\perp(\theta_i) = 0.5\,\rho_{\rm ISM}\,V_\star^2\,V_{\star \perp}(\theta_i)\, A_\perp(\theta_i)$, and $\Lambda_{\rm ff}$ and $\Lambda_{\rm tot}$ are the free--free and total cooling coefficients, respectively \citep{Myasnikov_1998}. On the other hand, the normalisation of the emission from the IL layer is determined by the volume of the cell:
\begin{equation}
    L_0 = \Delta V(\theta) = 2\,\pi\,R(\theta)\sin{(\theta)}\,H_{\rm IL}(\theta)\,\Delta l(\theta),
\end{equation}
where $\Delta l(\theta)$ is the length of the corresponding cell.

\subsection{Beyond the bow shock region}\label{subsec:Model_HII_region}

OB stars have strong UV radiation fields that ionise their surrounding media. If the star is supersonic, the size of the ionised region can be approximated as the size of the Str\"omgren sphere \citep{Tagle_1979}:
\begin{equation}
    R_{\rm Str} = 0.68\,{\rm pc}\,\left(\frac{Q_{\rm ion}}{10^{49}\,{\rm s^{-1}}}\right)^{1/3}\,\left(\frac{n}{10^3\,{\rm cm^{-3}}}\right)^{-2/3}\,\left(\frac{T}{10^4\,{\rm K}}\right)^{0.28},
    \label{Eq:R_Str}
\end{equation}
with $Q_{\rm ion}$ the ionising photon injection rate of the star and $n$ the numerical density of the neutral ISM \citep{Lacki_2014}.

In turn, a fraction of NT particles escape from the BS region without radiating significantly. These particles may diffuse and convect inside the ionised ISM, interacting with the local radiation and mass fields and producing multi-wavelength radiation. We focused on gamma rays and used a simplified one-zone model to estimate their luminosity, as this is the most conspicuous component of this emission.

Let $L_{\rm NT}$ and $L_{\rm BS,e(p)}$ be the luminosity injected into and radiated by electrons (protons) in the BS. The cosmic-ray luminosity injected in the ionised ISM is $L_{\rm HII,e(p)}$ = $L_{\rm NT,e(p)} - L_{\rm BS,e(p)}$. In turn, the fraction of this luminosity radiated by a certain radiation process can be estimated as
\begin{equation}
    L_{\rm rad,e(p)} \sim \frac{t_{\rm esc,e(p)}}{t_{\rm rad,e(p)}} L_{\rm HII,e(p)},
    \label{Eq:L_rad_Str}
\end{equation}
where $t_{\rm esc}$ and $t_{\rm rad}$ are the escape timescale and radiation process cooling timescale, respectively (formula valid for $t_{\rm esc} < t_{\rm rad}$). We calculated the former in terms of the convection and diffusion timescales, $t_{\rm esc}^{-1} = \left(t_{\rm conv}^{-1} + t_{\rm dif}^{-1}\right)^{-1}$. In particular, $t_{\rm conv} = R_{\rm HII}/V_\star$, and we adopted the diffusion coefficient proposed by \cite{Gabici_2006}, which is suitable for these structures:
\begin{equation}
    t_{\rm diff} = \frac{R_{\rm HII}^2}{6D_{\rm diff}}; \quad D_{\rm diff} = 3\times 10^{27} \chi \left(\frac{E/{\rm GeV}}{B/\left(3{\rm \mu G}\right)}\right)^{0.5}\; {\rm cm^2\,s^{-1}},
    \label{Eq:t_diff}
\end{equation}
with $\chi < 1$. We fixed $\chi = 0.1$ and $B = 10$~$\mu$G \citep{Gabici_2006}, which yields $D_{\rm diff}\approx 2\times 10^{26}$~cm$^2$~s$^{-1}$ at 1~GeV, adopted as the energy reference for both protons and electrons.

Relativistic particles can interact with the ambient nuclei from the ionised region leading to the emission of gamma-ray photons. For relativistic protons, these interactions are p-p collisions in which the resultant neutral pions decay into gamma rays, while relativistic electrons interact with the nuclei electrostatic field and emit NT bremsstrahlung. If $\mu$ is the mean molecular weight ($\mu\approx 0.6$ for a fully ionised gas with solar abundances), and $n_{\rm ISM}$ the number density inside the sphere, the ambient proton density is $n_p = \mu\,n_{\rm ISM}$. Then, the p-p and NT-bremsstrahlung timescales are \citep{Atoyan_1995,Aharonian_1996}
\begin{equation}
    t_{\rm p-p} \approx t_{\rm br} \approx 10^{15} \, \left( \frac{n_p}{\mathrm{cm}^{-3}}\right)^{-1}~\mathrm{s}.
    \label{Eq:t_pp}
\end{equation}

Approximately 15\% of the energy available in each p-p collision goes to gamma rays \citep{Kelner_2006}. Moreover, if the proton energy distribution also peaks at $E_p \sim 1$~GeV, as is the case here, the p-p SED peaks at $\epsilon \sim 100$~MeV. Taking into account the fact that ionisation losses are not dominant at energies $E \gtrsim 1$~GeV \citep{Atoyan_1995}, we approximated the gamma-ray luminosity as
\begin{equation}
    L_{\rm \epsilon,p-p}(\gtrsim 100~{\rm MeV}) \sim 0.1\;\left(\frac{t_{\rm esc}}{t_{\rm p-p}}\right)\;L_{\rm HII,p}.
    \label{Eq:L_pp}
\end{equation}
On the other hand, the NT-Bremsstrahlung SED peaks at $\epsilon \sim 1$~GeV. In this case, most of the radiation is emitted around that energy, so that we can approximate
\begin{equation}
    L_{\rm \epsilon,br}(\gtrsim 1~{\rm GeV}) \sim \left(\frac{t_{\rm esc}}{t_{\rm br}}\right)\;L_{\rm HII,e}.
    \label{Eq:L_Br}
\end{equation}
In addition, p-p collisions also lead to electron--positron pair production. These secondary electrons could also contribute to the NT-Bremsstrahlung SED. However, their luminosity would be approximately half that in p-p gamma rays, and thus the p-p gamma-ray luminosities estimated above are already correct within an order of magnitude. 


\section{Results and discussion}\label{sec:Results}


\subsection{Thermodynamics}\label{subsec:thermo}

\begin{figure*}
    \centering  
    \includegraphics[angle=270, width=0.49\textwidth]{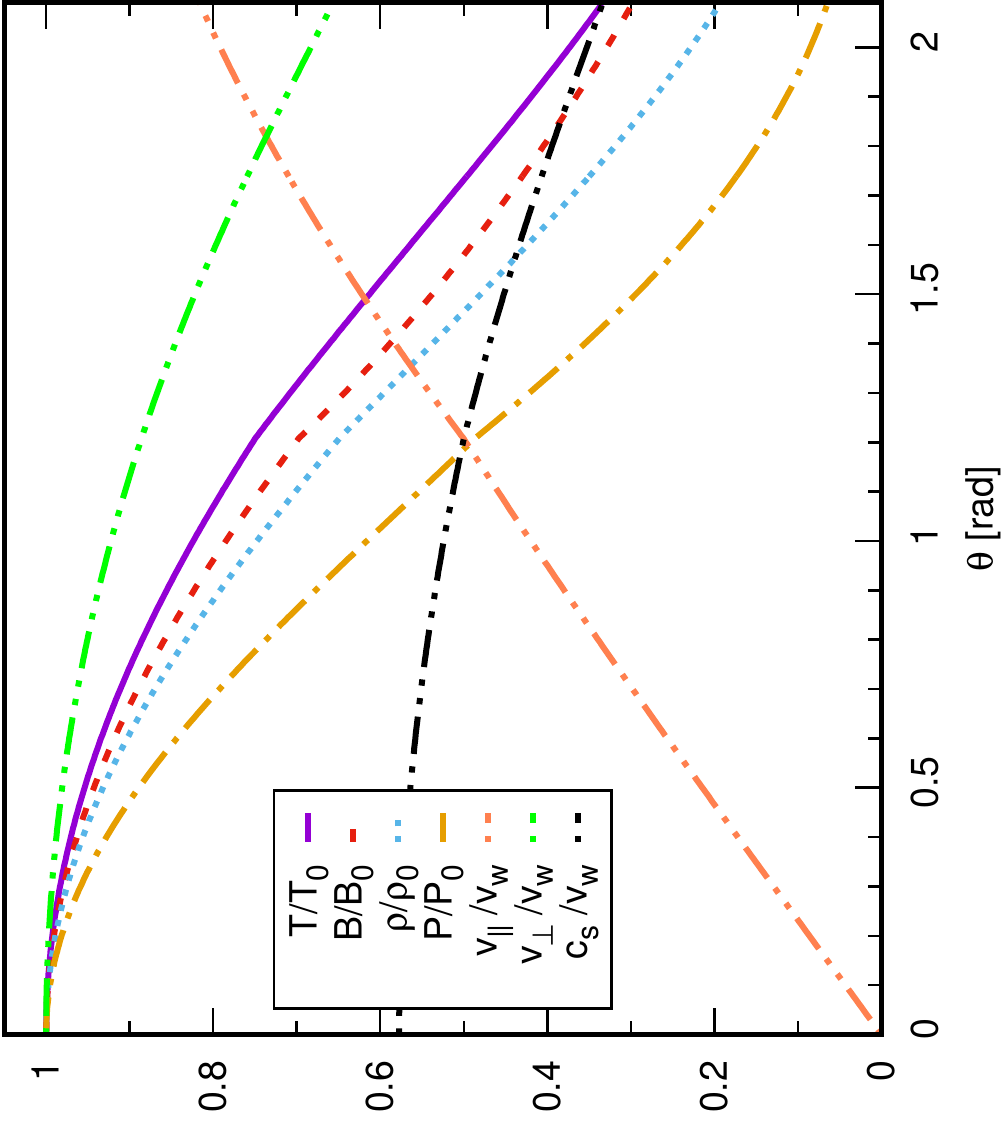}
    \includegraphics[angle=270, width=0.49\textwidth]{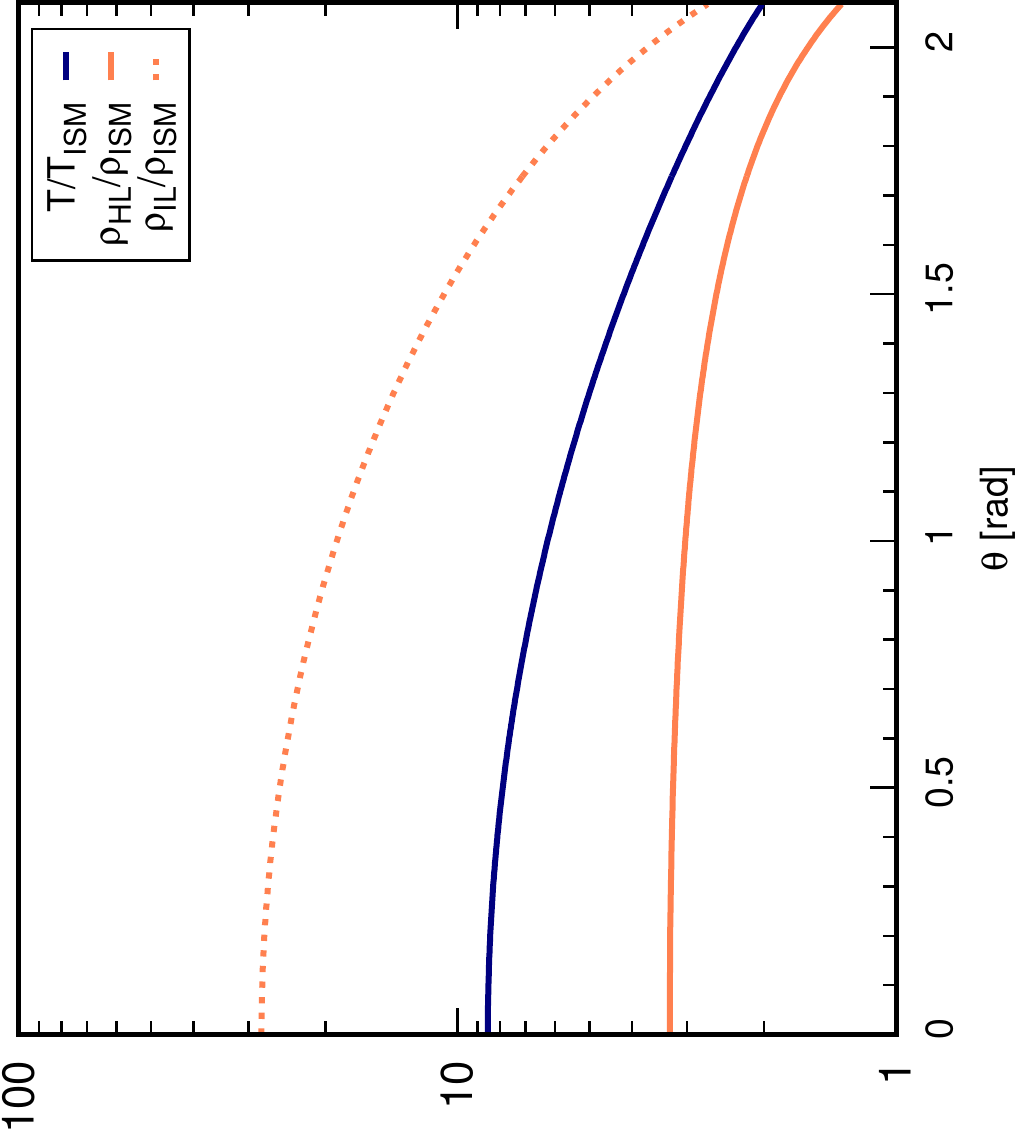}
    \caption{Thermodynamical quantities along the BS of \BD. {\it Left panel:} Thermodynamic variables in the RS as a function of $\theta$. The temperature, magnetic field, density, and pressure are normalised to their values at the apex ($\theta=0$), while the tangential, perpendicular, and sound velocities are normalised to the stellar wind velocity. {\it Right panel:}  Density of the IL (dotted coral line) and HL (solid coral line) and temperature of the HL (blue solid line), all as a function of the angle $\theta$; the variables are normalised to the un-shocked ISM values. }
    \label{fig:thermo}
\end{figure*}

As an example, in Fig.~\ref{fig:thermo} we show the most relevant thermodynamic quantities from the RS and the FS of the BS of \BD as a function of the angle $\theta$, up to the angle $\theta_{\rm max}$. In particular, we set $\theta_{\rm max}= 120\degree$ for \BD and \BN, $\theta_{\rm max}= 135\degree$ for Vela-X1, $\theta_{\rm max}= 80\degree$ for G1 and $\theta_{\rm max}= 90\degree$ for G3, using the criteria mentioned in Sect.~\ref{subsubsec:hydro_FS}.

Given that the FS of the BS of \BD is not hypersonic but $M(\theta) \gtrsim 1$, the resulting density in the HL (Eq.~\ref{Eq:rho_1}) is $\rho_{\rm HL}(\theta) \sim 3\,\rho_{\rm ISM}$, below typical values for strong shock conditions. The conversion of kinetic energy into thermal energy is more efficient for small values of $\theta$, and the shocked ISM reaches temperatures of $T_{\rm HL} \sim 70\,000~$K near the apex. Because of cooling, the shocked ISM material is compressed in the IL due to the corresponding temperature decrease, and the density reaches values above $10\,\rho_{\rm ISM}$ up to an angle $\theta \sim 85\degree$. 

\subsection{Radio emission}\label{subsec:radio_results}

The synchrotron radiation from the RS depends on the magnetic field, which is determined by the pressure in the subsonic region, and by the velocity and density of the fluid after the sonic point.  Figure~\ref{fig:thermo} shows that the pressure in the RS decays with $\theta$ as the shock becomes more oblique. As a result, the magnetic field also decays gradually, leading to stronger synchrotron emission near the apex. This is consistent with the morphology and brightness distribution seen in the resolved images of BSs \citep{Benaglia_2010, Moutzouri_2022}.

Regarding the FS, the free--free emission from both the HL and the IL depends on the density and temperature, as shown in Eqs.~(\ref{Eq:L_ff} -- \ref{Eq:L_ff_1}). In particular, the high temperature of the HL makes its free--free emission negligible at radio frequencies, as $\Lambda_{\rm ff}(T)/\Lambda_{\rm tot}(T) \ll 1$. On the other hand, the IL emission could be important at radio, although the layer (and therefore its emission) could be reduced significantly because of instabilities (see Sects.~\ref{subsec:BD43_results} and \ref{subsec:BD60_results} for further details).

\subsubsection{\BD}\label{subsec:BD43_results}

\begin{figure}
    \centering
    \includegraphics[width=\linewidth]{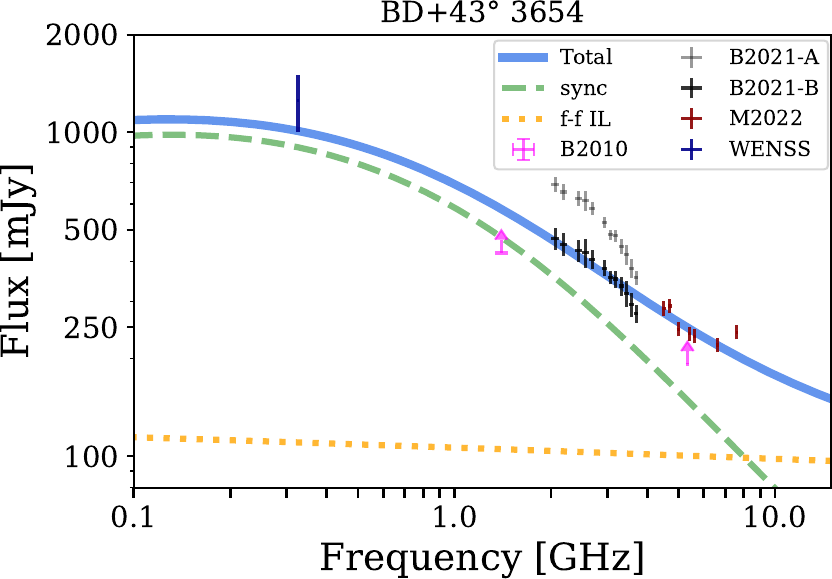}
    \caption{SED of the BS of \BD between 100~MHz and 15~GHz. The dashed green, dotted orange, and solid light blue lines are the synchrotron, free--free (IL) and total emission curves calculated with our model. The magenta arrows are the lower limits of \cite{Benaglia_2010}. The grey data points show the SED reported by \cite{Benaglia_2021} when selecting contours > 1.5~\mJybeam, and the black data points show the SED from contours > 2.3~\mJybeam. The dark red data points show the SED reported by \cite{Moutzouri_2022}. The dark blue bar is the total flux obtained from the WENSS emission map at 325~MHz (Appendix~\ref{appendix:radio_data}).}
    \label{fig:SED_BD+43}
\end{figure}

In Fig.~\ref{fig:SED_BD+43} we show the radio SED of the BS of \BD together with the data from \cite{Benaglia_2010}, \cite{Benaglia_2021}, and \cite{Moutzouri_2022} and our estimate from Appendix~\ref{appendix:radio_data}. In particular, we considered the VLA observations from \cite{Benaglia_2010} as lower limits, as they were not sensitive enough to reach the fainter emission from the more extended regions of the BS and thus underestimated the total integrated fluxes. We also show two SEDs reported by \cite{Benaglia_2021}. One of them integrated the flux densities within contours above 1.5~mJy beam$^{-1}$, while the other considered contours above 2.3~mJy beam$^{-1}$; hereafter we refer to these datasets as B2021--A and B2021--B, respectively. The contours selected in B2021--A may include regions that do not correspond to the BS, as inferred from infrared emission maps from \textit{WISE} \citep[see Fig.~1 in][]{Moutzouri_2022}. We thus considered that the reported fluxes in B2021--A could be overestimated, and those of B2021--B are more representative of the true spectrum of the BS.

We find that the predicted free--free emission for $\eta_{\rm H} = 1$ is too high: it overestimates the radio fluxes by a factor of $\sim 2$ and leads to a thermal-dominated spectrum, in discordance with the observed spectral indices. We thus imposed $\eta_{\rm H} < 0.3$ to be consistent with the available data, meaning that the IL is significantly reduced. Kelvin Helmholtz instabilities can justify this drastic reduction in the IL due to the difference between the RS and the FS shocked flow velocities \citep{Comeron_Kaper_1998}. However, characterising the phenomena that produce these effects is beyond the scope of this paper.

Just as a reference, we considered that $\sim 15$\% of the injected power in the RS goes to cosmic rays (that is, $f_\mathrm{NT}\sim 0.15$) and, to explain observations, that 10\% of that power must go to electrons (leading to $f_\mathrm{NT,e}\sim 0.015$). In addition, we fixed the magnetic field setting $\eta_{\rm B} \sim 0.2$; the adopted parameters are summarised in Table~\ref{table:parameters}. The value of $\eta_{\rm B} \sim 0.2$ would be consistent with the adiabatic compression of magnetic field lines from the stellar wind for a surface stellar magnetic field of $B_\star \sim 300$~G \citep{del_Palacio_2018}. 
We note that slightly different choices of parameters can also lead to a similar spectrum, but the overall conclusions would not be affected and thus, for simplicity, we took the aforementioned values as representative.

The luminosity injected into cosmic rays is at least $L_{\rm NT} \sim 10^{35}~$erg\,s$^{-1}$, in the form of NT electrons, which is $\sim 1.5$\% and $\sim 0.8$\% of the shocked and the total wind luminosity, $L_{\rm w} = 0.5\,\dot{M}\,v_{\rm w}^2$, respectively. However, the energy in hadronic cosmic rays may be potentially much higher, so $L_{\rm NT}\gtrsim 10^{36}~$erg\,s$^{-1}$ are plausible. The magnetic field results in $B \sim 30~\mu$G near the apex, and the ratio between the magnetic and NT electron energy densities is $U_{\rm B}/U_{\rm NT,e} \sim 4$, above (but not very far from) equipartition of $B$ with NT electrons. Regarding the maximum energies, both electrons and protons reach energies of $E_{\rm max} \lesssim 100$~TeV near the apex assuming Bohm diffusion. In addition, our model predicts radio fluxes consistent with B2021--B, whereas reaching the flux densities from B2021--A would require even more extreme parameters, which also supports the choice of B2021--B. Finally, we also verified that the predicted NT emission in X-rays (mostly synchrotron) between 0.4 and 4~keV is $L_{\rm X} = 2\,\times\,10^{30}~$erg\,s$^{-1}$, consistent with the upper limit of $L_{\rm X} < 7.6\,\times\,10^{30}~$erg\,s$^{-1}$ \citep{Toala2016}. 
 
\begin{figure}[ht]
    \centering
    \includegraphics[angle=270,width=\linewidth]{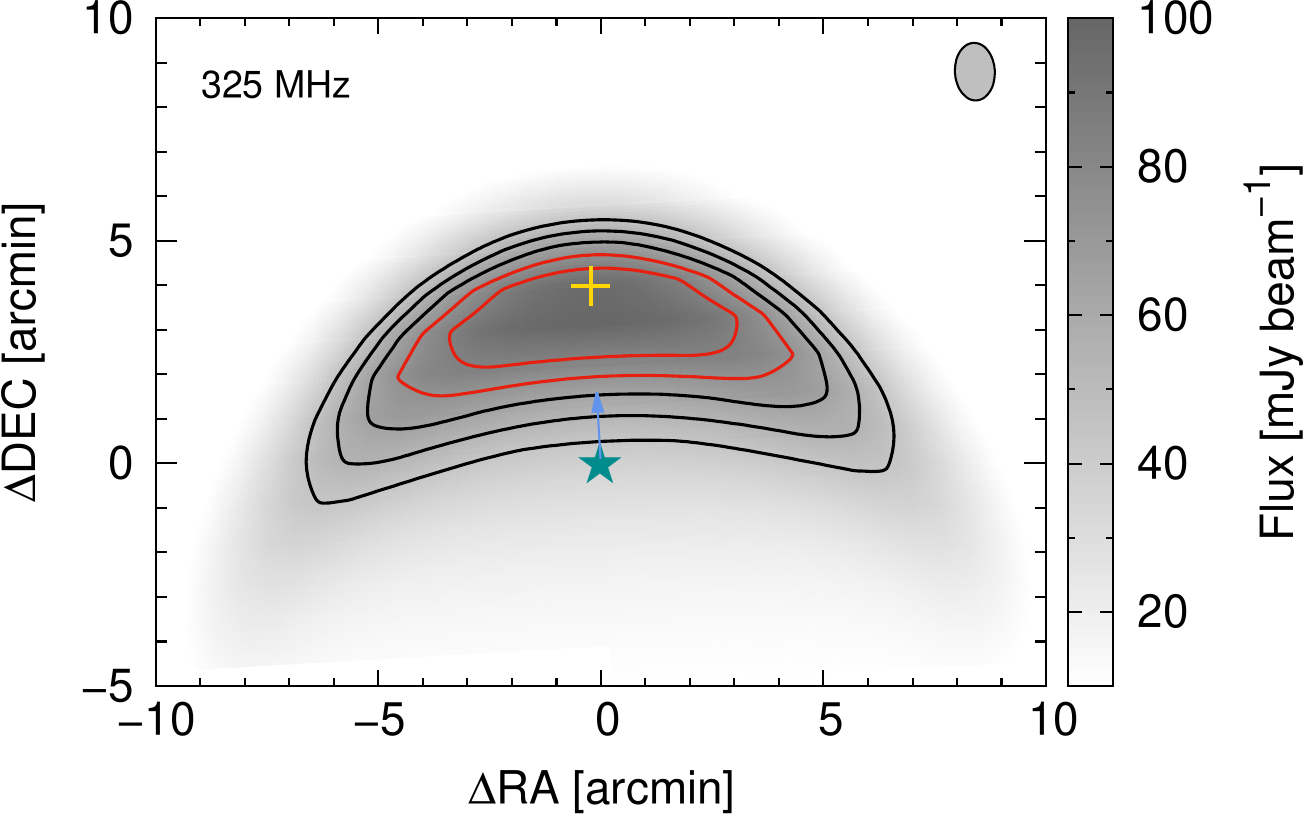}\\
    \includegraphics[angle=270,width=0.95\linewidth]{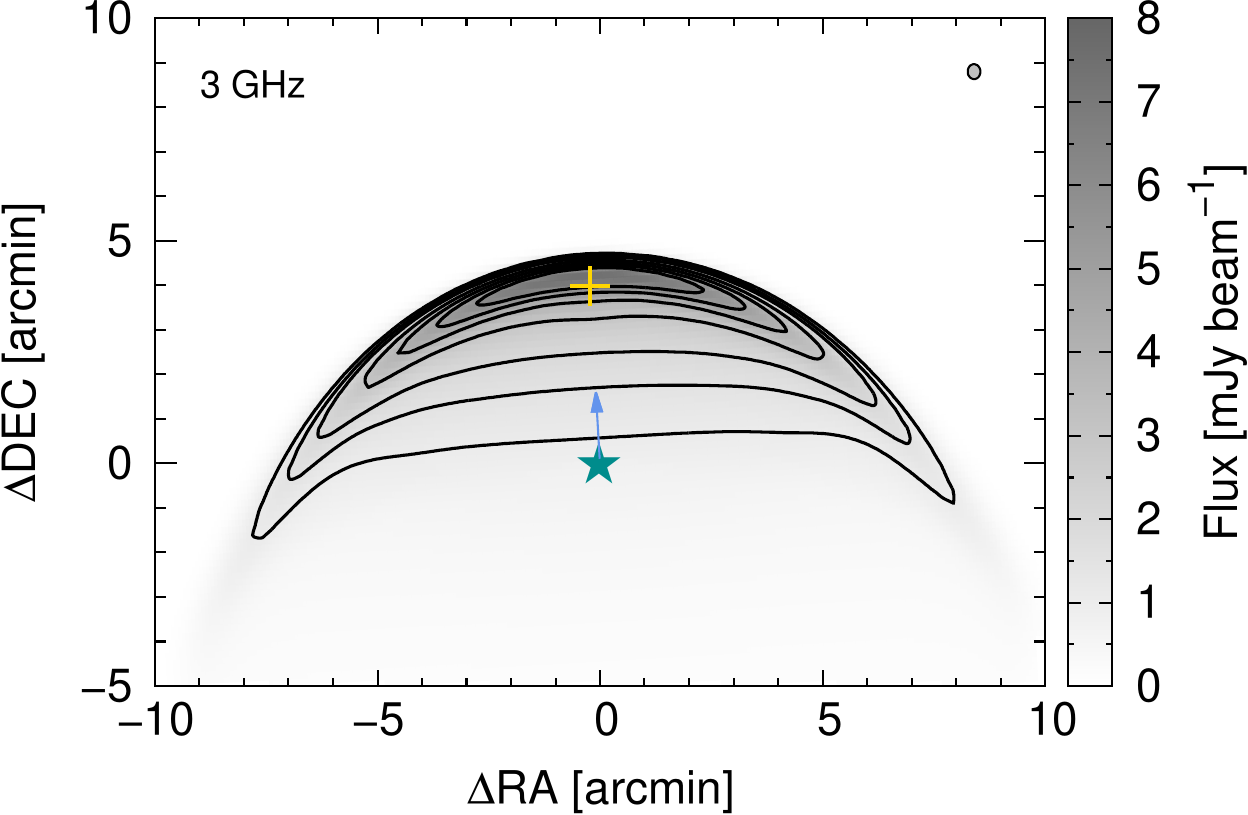}\\
    \includegraphics[angle=270,width=0.95\linewidth]{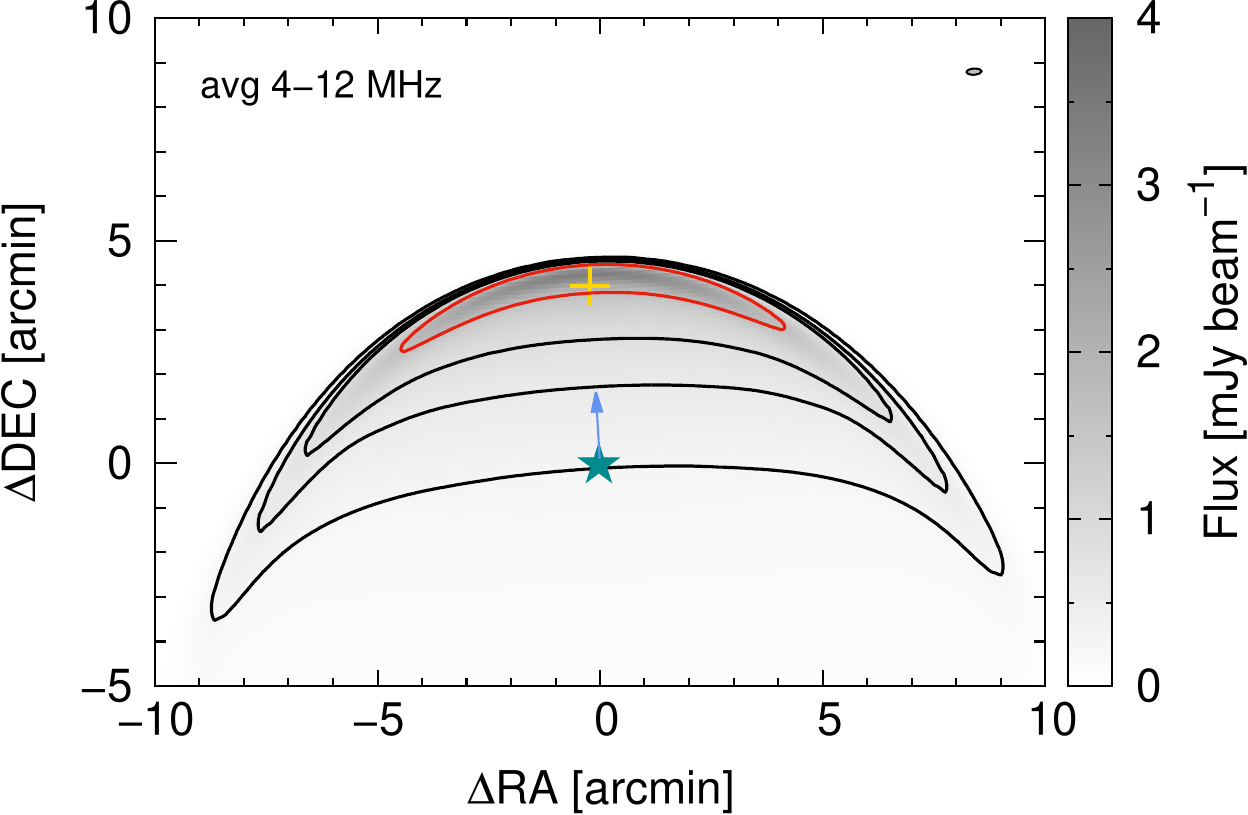}
    \caption{Simulated emission maps of the BS of \BD. Black contours match the observed ones, while red contours are above them. {\it Top panel:} Map at 325~MHz with a beam size of $54" \times 77"$. Black contour levels are 44, 54, and 64~\mJybeam; red contour levels are 74 and 84~\mJybeam. {\it Middle panel:} Map at 3~GHz with a beam size of $20.2"\times 12.5"$. Contour levels are 1, 1.5, 2, 3, 4, 5, and 6~\mJybeam. {\it Bottom panel:} Average map at 4--12~GHz with a beam size of $20"\times10"$. Black contour levels are 0.3, 0.6, and 0.9~\mJybeam; the red contour level is 1.8~\mJybeam.}
    \label{fig:maps_BD+43}
\end{figure}

\cite{Brookes_2016}, \cite{Benaglia_2021}, and \cite{Moutzouri_2022} reported emission maps of \BD at 325~MHz, 3~GHz, and 4--12~GHz, respectively. In Fig.~\ref{fig:maps_BD+43} we show our synthetic maps at those same frequencies. Our model successfully reproduces the spatial morphology observed on a qualitative level: the BS is brighter at the apex, where the shock is stronger and the magnetic field is $B \sim 30~{\rm \mu G}$, and the angular extension also matches the observed one. On a more quantitative level, the contour levels of the synthetic map at 3~GHz follow the same emission contours as the observed map, while the maps at 325~MHz and 4--12~GHz slightly overestimate the surface brightness by factors of $\sim 1.5$ and $\sim 2$, respectively. Nevertheless, the uncertainty on $V_{\star,r}$ can also affect these results, as larger values of $V_{\star,{\rm r}}$ lead to a more extended BS in the plane of the sky, diluting the surface brightness.

\begin{figure}[ht]
    \centering
    \includegraphics[angle=270,width=\linewidth]{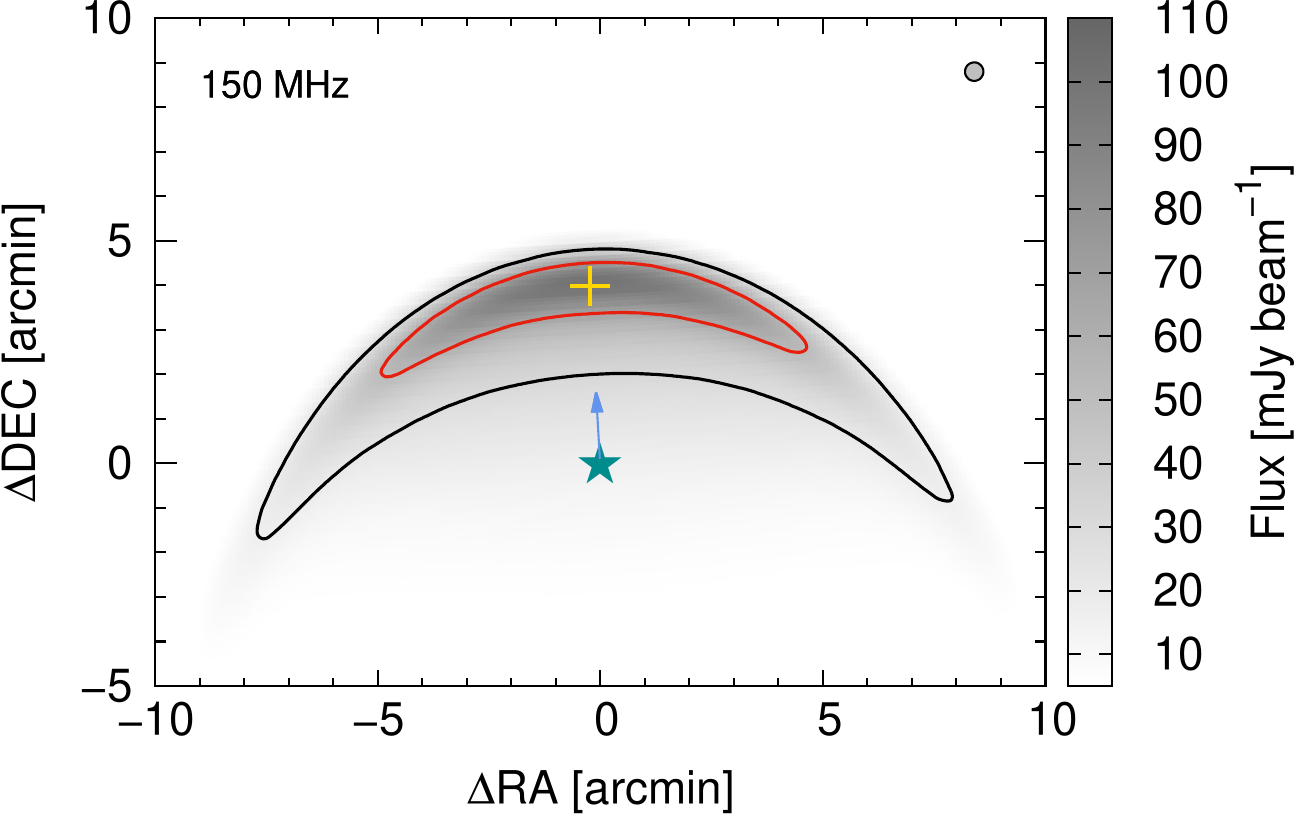}\\    \includegraphics[angle=270,width=\linewidth]{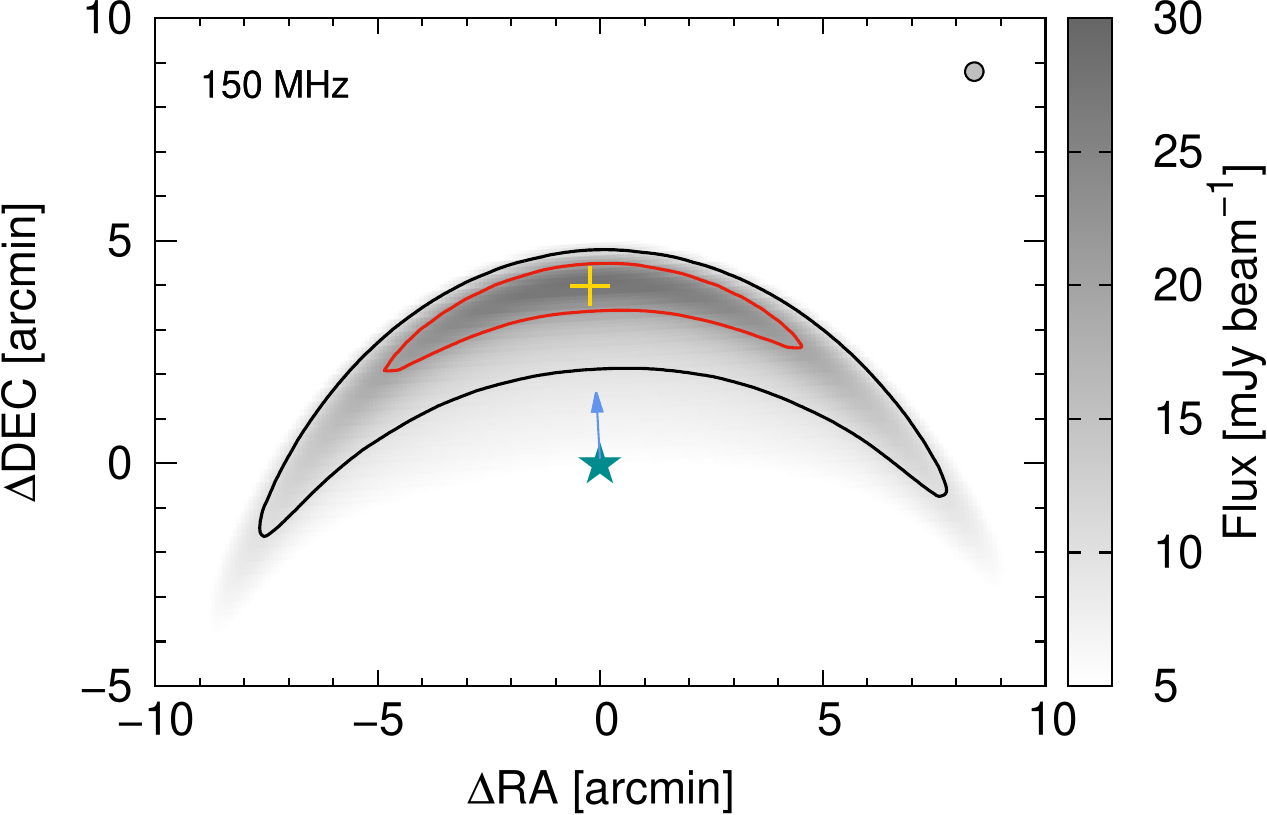}
    \caption{Simulated emission maps of the BS of \BD at 150~MHz using the single index injection ({\it top panel}) and the piecewise injection ({\it bottom panel}). The beam size is $25" \times 25"$. The black contour is 10~\mJybeam (similar to the rms of the map from the TGSS survey; see Fig.~\ref{fig:TGSS_BD43}), while the red contour levels are 20, 40, 80, and 120~\mJybeam.}
    \label{fig:maps_150_BD+43}
\end{figure}

Finally, we produced synthetic emission maps at 150~MHz using two different injection prescriptions. In Fig.~\ref{fig:maps_150_BD+43} (top) we use an injection function with a single spectral index $p=-2.55$, while in Fig.~\ref{fig:maps_150_BD+43} (bottom) we use the piecewise function from Eq.~(\ref{Eq:Q_inj}). The single index prescription overestimates the fluxes at 150~MHz by a factor of $\sim 7$ (see Appendix~\ref{appendix:radio_data}). On the other hand, the piecewise injection is consistent with the non-detection of the BS above (3-$\sigma$) noise levels of $\gtrsim20$~\mJybeam. In addition, using the single index prescription requires fixing $\eta_{\rm B} \approx 0.4$ and that $\sim 40\%$ of $\Delta L_\perp$ goes to electrons to fit the spectrum between 1 and 12~GHz. Given that these assumptions are very extreme, and that the single index prescription largely overestimates the radiation at 150~MHz, we conclude that a piecewise injection function is needed.

\subsubsection{\BN}\label{subsec:BD60_results}

The spectral index derived by \cite{Moutzouri_2022} at 4--12~GHz also implies a steep energy distribution of relativistic electrons with energies from $\sim 1$~GeV to at least $\sim$10~GeV. To recover this steep radio spectrum requires a reduction in the IL ($\eta_{\rm H}<0.15$). Moreover, the available luminosity in the RS is insufficient to explain the observed synchrotron SED with the single index prescription, so the electron distribution should be much harder below $\sim 1$~GeV. We thus adopted the injection spectrum from Eq.~(\ref{Eq:Q_inj}), assumed that $\sim 5$\% of the kinetic energy is converted into electrons, and set the reference value $\eta_B \approx 0.2$, which yields a magnetic field of  $50~\mu$G near the apex.

\begin{figure}[ht]
    \centering
    \includegraphics[angle=270,width=0.49\textwidth]{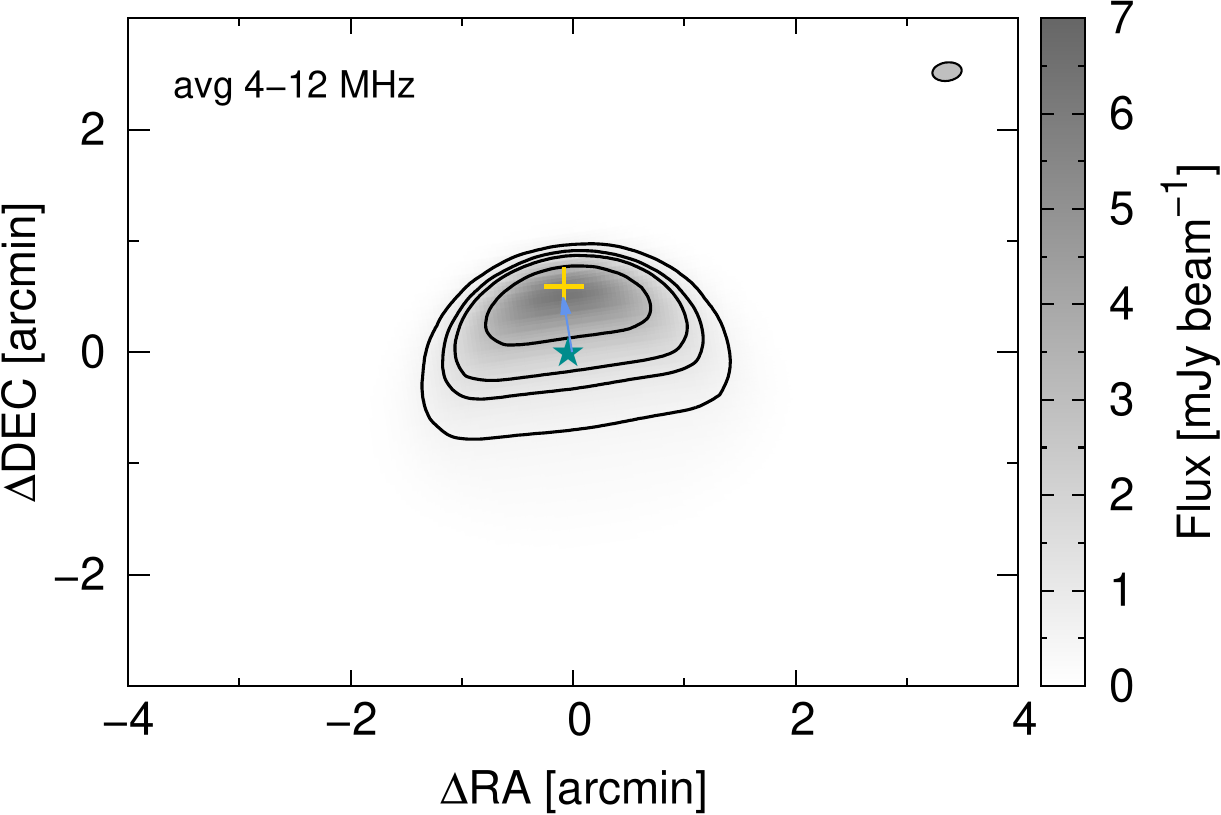}
    \includegraphics[angle=270,width=0.48\textwidth]{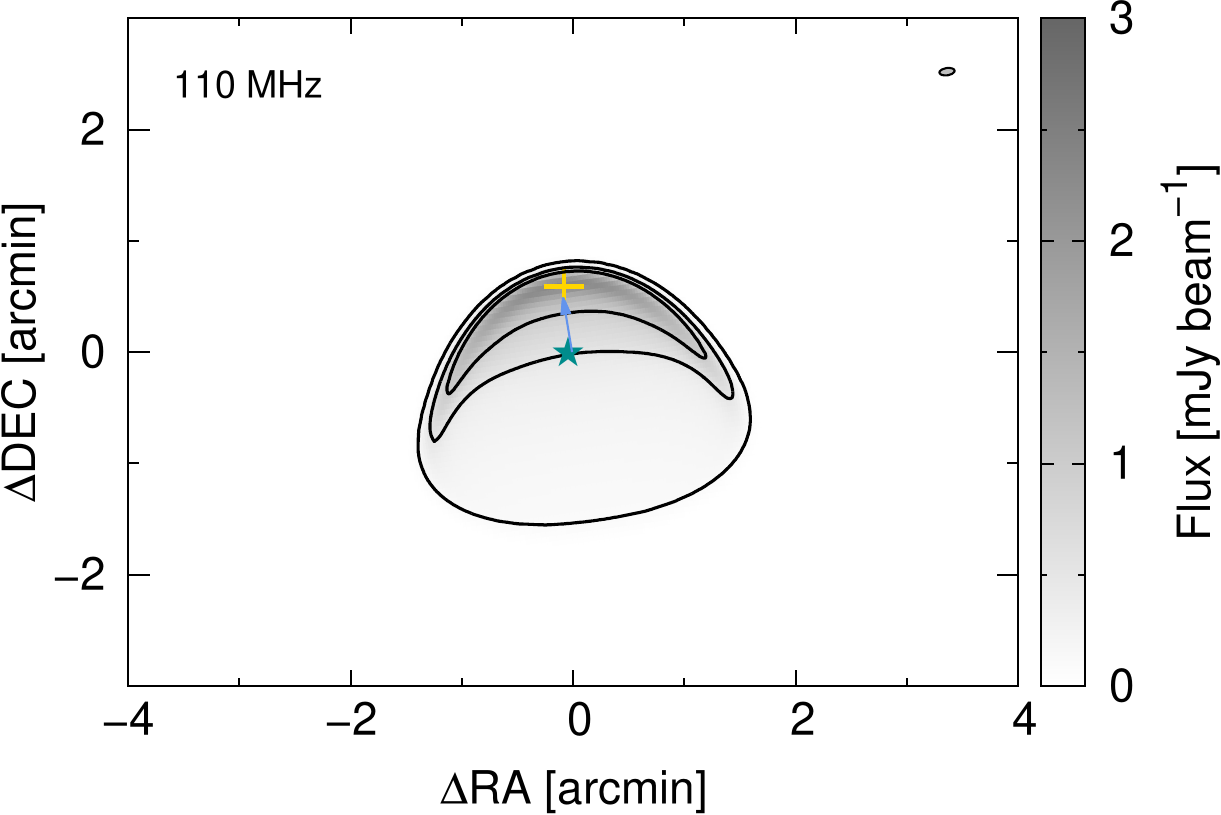}
    \caption{Simulated emission maps of the BS of \BN. {\it Top panel:} Average map between 4 and 12~GHz. The contour levels are 0.5, 1, 1.5, and 3 mJy beam$^{-1}$. {\it Bottom panel:} Map at 110~MHz. The beam size is $4" \times 4"$, and the contour levels are 0.1, 0.5, and 1~\mJybeam.}
    \label{fig:map_avg_BD+60}
\end{figure}

In Fig.~\ref{fig:map_avg_BD+60} we show two synthetic emission maps of the BS of \BN. In Fig.~\ref{fig:map_avg_BD+60} (top) we show the average emission between 4 and 12 GHz, while in Fig.~\ref{fig:map_avg_BD+60} (bottom) we predict the observed emission at 110~MHz. We note that our modelled map is in good accordance with the flux density of $\sim 6$~\mJybeam near the apex of the BS seen in Fig.~1 of \cite{Moutzouri_2022}. However, we highlight that the total luminosity reported by \cite{Moutzouri_2022} is most likely to be highly overestimated due to the presence of dense, ionised clumps to the west of \BN \citep{Moore2002}. These clumps emit bright thermal emission that dominates the single-dish observations between 4 and 12 GHz.

We also predict integrated fluxes of $\sim110$~mJy at 150~MHz and $\sim120$~mJy at 325~MHz from the BS of \BN. However, these values depend strongly on the assumed electron energy distribution below $\sim1$~GeV, which can only be inferred directly by measuring their synchrotron emission at $\nu \sim 100$~MHz. Hence, low-frequency, high-angular-resolution observations would allow us to better constrain the injection spectrum, and therefore the particle acceleration processes in the BS.

\subsection{G1, G3, and Vela X-1}

\begin{figure}[t]
    \centering
    \includegraphics[width=\linewidth]{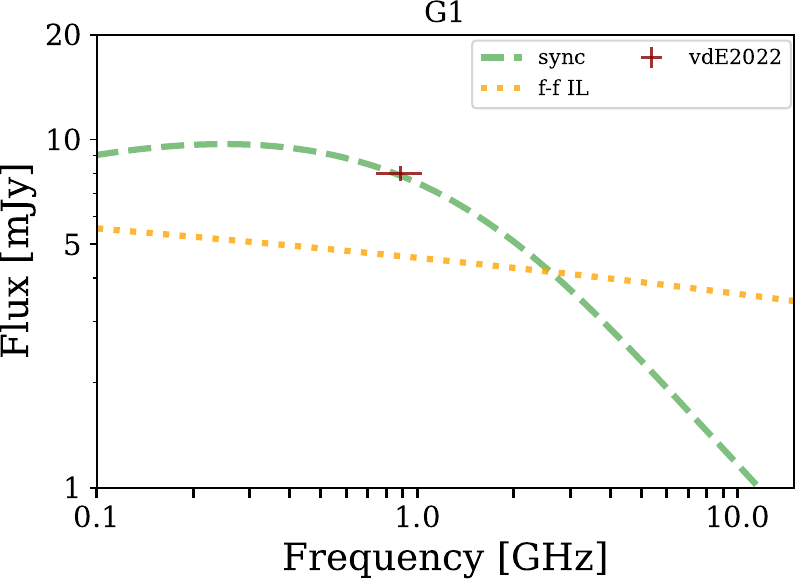} \\[2pt]
    \includegraphics[width=\linewidth]{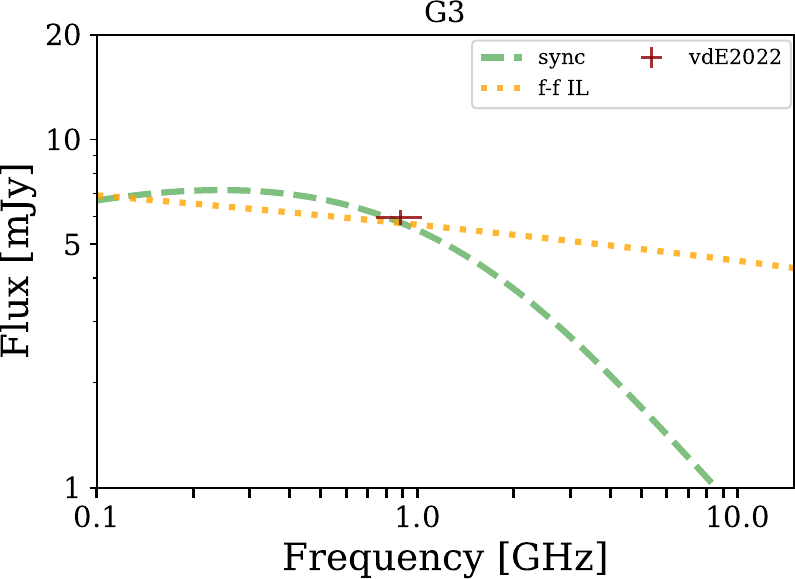} \\[2pt]
    \includegraphics[width=\linewidth]{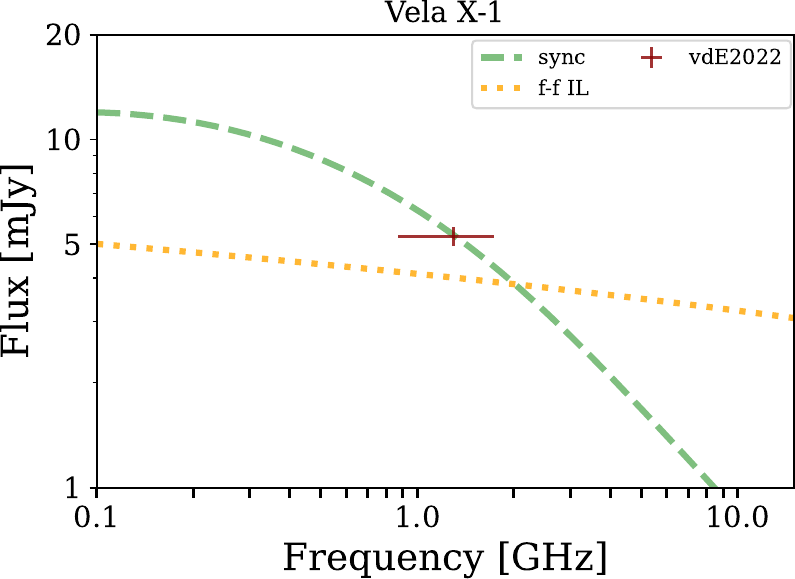}
    \caption{Modelled SEDs between 100 MHz and 15 GHz for G1 (top), G3 (middle), and Vela X-1 (bottom panel). In all cases, we consider two different scenarios: a NT one with a synchrotron spectrum (dashed green line) and a thermal one with free-free emission (dotted orange line). The red cross is the flux reported by \cite{VdE_2022} for G1 and G3, and by \cite{VdE2022_VelaX1} for Vela X-1.}
    \label{fig:SED_G1-G2-Vela}
\end{figure}

In Fig.~\ref{fig:SED_G1-G2-Vela} we show the SEDs for the sample of BSs reported by \cite{VdE_2022} and \cite{VdE2022_VelaX1}. For the case of G1, the flux density observed at 887.5~MHz cannot be explained solely as thermal emission, as even considering $\eta_\mathrm{H}=1$ yields a flux density of $\approx 5$~mJy and the observed flux is $\approx 9$~mJy. Moreover, the better studied BSs seem to suggest values of $\eta_\mathrm{H} < 0.3$. On the other hand, a purely NT scenario requires a high --but not unrealistic-- conversion of kinetic energy into NT energy. As a reference case, we can explain the observed flux density with a significant synchrotron contribution adopting $\eta_\mathrm{H} = 0.1$, $\eta_{\rm B} = 0.2$ and $f_{{\rm NT,e}} = 0.03$. However, it is not currently possible to determine which component is dominant in this system, as hybrid scenarios are also plausible.

For the case of G3, assuming $\eta_H = 1$ overestimates the fluxes by a factor of $\sim 2$. Thus, the emission can be of thermal origin if the IL width parameter is $\eta_H \sim 0.45$. Alternatively, we can explain the spectrum with reasonable NT parameters by setting $\eta_{\rm B} = 0.2$ as a reference value and assuming that $f_{{\rm NT,e}} = 0.01$. Although G1 and G3 have similar parameters, the reported mass loss rate of the latter is almost twice as large, favouring a NT-dominated scenario \citep[as the synchrotron luminosity scales roughly as $\propto \dot{M}_\mathrm{w}^{1.5}$;][]{del_Palacio_2018}. The similarities between the parameters assumed for \BD and \BN, and the NT scenario of G1 and G3, could suggest that indeed in G1 and G3 the radio spectrum can be dominated by synchrotron emission and that the IL is significantly evaporated. 
We note that this interpretation is in tension with that of \cite{VdE_2022}, who favoured a thermally dominated scenario for the systems G1 and G3. Nonetheless, our more detailed model considers an injection function of NT particles that relaxes the energy budget in a NT scenario and reconciles such a scenario with the observations better. Unfortunately, the large uncertainties in the system parameters together with the lack of a good spectral coverage of G1 and G3 prevent us from firmly concluding on the nature of their radio emission. 

Finally, for Vela X-1 we obtain that a NT-dominated scenario can explain the observed radio flux density for reasonable parameters, but the BS cannot be solely thermal. Considering $\eta_{\rm H} = 1$ yields a free-free flux density of $\approx 4$~mJy at 887.5 MHz, slightly below the $\approx 6$~mJy observed. In the NT scenario, we assumed that $\sim1\%$ of the available kinetic energy in the RS shock goes to electrons, while we set $\eta_{\rm B} = 0.2$. This leads to a magnetic-to-NT electron energy density ratio of $U_{\rm B}/U_{\rm NT,e} \sim 5$, and an upper limit for the IL width of $\eta_{\rm H} < 0.3$. Moreover, a NT-dominated scenario is supported by analysis of the observed H$_\alpha$ emission from the BS.
We can explain the peak surface brightness of H$_\alpha$ reported by \cite{Gvaramadze_2018} by setting $\eta_{\rm H} = 0.1$ (see Appendix~\ref{appendix:H_alpha}). This result strongly encourages the interpretation of a synchrotron-dominated spectrum in Vela X-1 and the significant evaporation of the IL. We note that \cite{VdE2022_VelaX1} favoured a thermal scenario for Vela X-1 based on matching the peak brightness in H$_\alpha$ using a one-zone model. Nonetheless, one-zone models are not as adequate as multi-zone models to interpret information from emission maps. 


\begin{table*}[t]
    \centering
    \renewcommand{\arraystretch}{1.3} 
    \caption[]{Main results regarding the radio emission from the modelled stellar BSs. Column 3 shows the power injected in the RS by the stellar wind typically, $L_{\rm w\perp} \approx 0.5 - 0.7 L_{\rm w}$). Columns 4--6 are model parameters: the fraction of the injected power that goes into electrons, the electrons-to-magnetic energy density ratio, and the IL width parameter; we fix $\eta_{\rm B} = 0.2$ for all systems. Column 7 summarises our conclusions about the origin of the emission of each source.}
    \begin{tabularx}{\linewidth}{l c | c | c c c | X}
    \hline\hline\noalign{\smallskip}
    System & Radiation type & $L_{\rm w\perp}$ [erg s$^{-1}$]  & $f_{\rm NT, e}$        & $U_{\rm B}/U_{\rm NT,e}$   & $\eta_{\rm H}$    & Comments \\ 
    \hline\hline
    \BD     & NT    & $\approx 7\times 10^{36}$  & 0.013 & $\sim 4$& < 0.3   & The observed spectral index at $\nu > 1$~GHz implies a NT spectrum.\\
    \hline
    \BN     & NT    & $\approx 1\times 10^{36}$  & 0.04  & $\sim 1.5$ & < 0.15  & The observed spectral index at $\nu > 1$~GHz implies a NT spectrum.\\
    \hline
    G1      & Hybrid/NT & $\approx 1\times 10^{35}$ & 0.03  & $\sim 2$ & 0.1     & The observed flux density at 887.5~MHz cannot be solely thermal and is most likely dominated by NT emission. \\
    \hline
    \multirow{2}{*}{G3}  & NT  & \multirow{2}{*}{$\approx 2\times 10^{35}$}  & 0.01 & $\sim 6$ & < 0.1 & The nature of the emission is inconclusive due to uncertainties in $\dot{M}_\mathrm{w}$ and the lack of broadband spectral coverage. \\[-3ex]
                        & Thermal &                                        & < 0.002 & > 30     & 0.45  &      \\
    \hline
    \multirow{2}{*}{Vela X-1} & NT  & \multirow{2}{*}{$\approx 1\times 10^{35}$}  & 0.013 & $\sim 5$ & < 0.3  & A reduction in the IL is needed to reproduce the H$_\alpha$ surface brightness, and the observed flux density at 887.5~MHz cannot be solely thermal. A NT-dominated spectrum is thus favoured. \\[-3ex]
                            & Thermal &                                        & < 0.003 & > 25       & 1  &   \\
    \bottomrule
    \end{tabularx}
\label{table:parameters}
\end{table*}

\subsection{Gamma-ray emission}\label{subsec:gamma_results}

We applied the prescription described in Sect.~\ref{subsec:Model_HII_region} to the confirmed NT BSs: \BD and \BN. In both systems, the behaviour of NT particles along the BS is dominated by advection. In consequence, $L_{\rm BS} \ll L_{\rm NT}$ and particles escape from the BS without cooling significantly, and thus $L_{\rm HII} \approx L_{\rm NT}$.

According to Eq.~(\ref{Eq:R_Str}), the HII region around \BD has a radius of $\sim 20$~pc. For the value of $D_{\rm diff}\approx 2\times 10^{26}$~cm$^2$~s$^{-1}$  obtained above (note that Bohm diffusion yields $D_{\rm diff}\sim 6 \times 10^{22}~$cm$^2$ s$^{-1}$), diffusion dominates the escape, with a timescale of $t_{\rm diff}\approx 3\times 10^{12}$~s. On the other hand, p-p and NT-bremsstrahlung timescales are much greater: $t_{\rm p-p} \sim t_{\rm br} \sim 3\times 10^{14}$. Then, particles may also diffuse from the HII sphere without radiating a significant fraction of their energy as high-energy photons. We estimate luminosities $L_{\rm \epsilon, p-p}(\gtrsim 100\,{\rm MeV}) \sim L_{\rm \epsilon, br}(\gtrsim 1\,{\rm GeV}) \sim 10^{33}$~erg\,s$^{-1}$, if we assume $f_{\rm NT,e}= 0.1\,f_{\rm NT,p}\approx f_{\rm NT}$. According to our model, the emission from the HII region could dominate the spectrum at gamma rays, as the IC luminosity from the BS at 1~GeV would be $L_{\rm \epsilon, IC}(\gtrsim 1\,{\rm GeV}) \sim 10^{31}$~erg\,s$^{-1}$. Additionally, the IC emission from the HII region with the cosmic microwave background and the star photon fields are negligible at gamma rays, as the SED from the former peaks at lower energies, while the latter is $L_{\rm IC} \propto U_\star \propto R^{-2}$. Lastly, we notice that the derived 0.1--1~GeV luminosities, reachable already only with primary electrons, are close to the upper limits found by \cite{Schulz_2014} (also $\sim 10^{33}$~\ergs), so the values taken above seem to suggest that \BD may be detectable in the near future. Alternatively, a lack of detection would provide constraints on the adopted parameters, in particular on the diffusion coefficient, which is the most uncertain of all.

Our model estimates a Str\"omgren sphere of $\sim 4$~pc for \BN. This smaller value is because of the much denser ISM and the later spectral type of the star compared with \BD. Again, diffusion dominates the timescales, being $t_{\rm diff} \approx 10^{11}$~s, while $t_{\rm p-p} \sim t_{\rm br} \sim 8\times 10^{13}$~s. Adopting $f_{\rm NT,e}=0.2\,f_{\rm NT,p}\approx f_{\rm NT}$ for this source, we estimate that the luminosity injected into NT particles in the BS is $L_{\rm NT} \sim 2\times 10^{35}$~erg\,s$^{-1}$. Taking into account this and the cooling timescales, we estimate gamma-ray luminosities emitted in the HII region around \BN of $L_{\rm \epsilon, p-p}(\gtrsim 100\,{\rm MeV}) \sim 10^{31}$, and $L_{\rm \epsilon, br}(\gtrsim 1\,{\rm GeV}) \sim 10^{32}$~erg\,s$^{-1}$ from primary electrons. 

Finally, regarding G1, G3 and Vela X-1, the energetics of their BSs make the systems undetectable at gamma rays with current instruments. In conclusion, our model predicts that the HII region around \BD could be detectable with the {\it Fermi}-Large Area Telescope (LAT). It is worth adding that the steepness of the electron energy distribution for \BD makes this source difficult to detect above 100~GeV \citep[see][for upper limits of other BSs]{HESS18}.

\section{Conclusions} \label{sec:conclusions}

We have studied the hydrodynamics and radiation from stellar BSs by developing a multi-zone model that takes both thermal and NT radiation into consideration. We applied this model to the sample of BSs that have been detected at radio frequencies to determine the nature of their emission. We find that stellar BSs can be efficient particle accelerators, potentially accelerating electrons and potentially protons up to energies $\lesssim 100$~TeV, and capable of transferring $\sim 1$--5\% of the shocked wind kinetic power to relativistic electrons (and an unknown amount to hadronic cosmic rays). After comparing our results with the observed spectra and emission maps, we conclude that the emission of the confirmed NT BSs (the BSs of \BD and \BN) can be explained if NT electrons follow a hard energy distribution below $\lesssim$2~GeV.

Regarding Vela X-1, a NT scenario is also favoured in view of the consistency with the H$_\alpha$ observations, which require a similar reduction in the IL as that inferred for \BD and \BN. Extending this result to the systems G1 and G3 would suggest that their radio spectra are also likely to be NT-dominated. However, the lack of wide spectral coverage, added to the uncertainties on the velocity of the stars and the stellar wind parameters, prevents us from drawing unambiguous conclusions. In order to provide better insights into the particle acceleration process and the nature of the radiation of these sources, observations at $\sim100$~MHz are needed.

Lastly, we give order-of-magnitude estimations of the gamma-ray emission of the HII regions around \BD and \BN. In particular, according to our simplified model, the HII region around \BD seems to be a good candidate to be eventually detected by {\it Fermi}-LAT. On the other hand, a great fraction of electrons (and protons) should escape, so the source should be injecting cosmic rays into our Galaxy with a luminosity from $L_{\rm CR} \sim 10^{35}$~erg s$^{-1}$ (at least for electrons) to perhaps $\gtrsim 10^{36}$~erg s$^{-1}$ (mostly protons). 


\begin{acknowledgements}
J.R.M. acknowledges support by PIP 2021-0554 (CONICET). The paper also received financial support from the State Agency for Research of the Spanish Ministry of Science and Innovation under grants 
PID2019-105510GB-C31/AEI/10.13039/501100011033/ 
and PID2022-136828NB-C41/AEI/10.13039/501100011033/, and by "ERDF A way of making Europe" (EU),
and through the ''Unit of Excellence Mar\'ia de Maeztu 2020-2023'' award to the Institute of Cosmos Sciences (CEX2019-000918-M). V.B-R. is Correspondent Researcher of CONICET, Argentina, at the IAR. This work was carried out in the framework of the PANTERA-Stars\footnote{\url{https://www.astro.uliege.be/~debecker/pantera/}} initiative.
\end{acknowledgements}

%
%




\appendix

\section{Additional radio data} \label{appendix:radio_data}
\begin{figure}
    \centering
    \includegraphics[width=\linewidth]{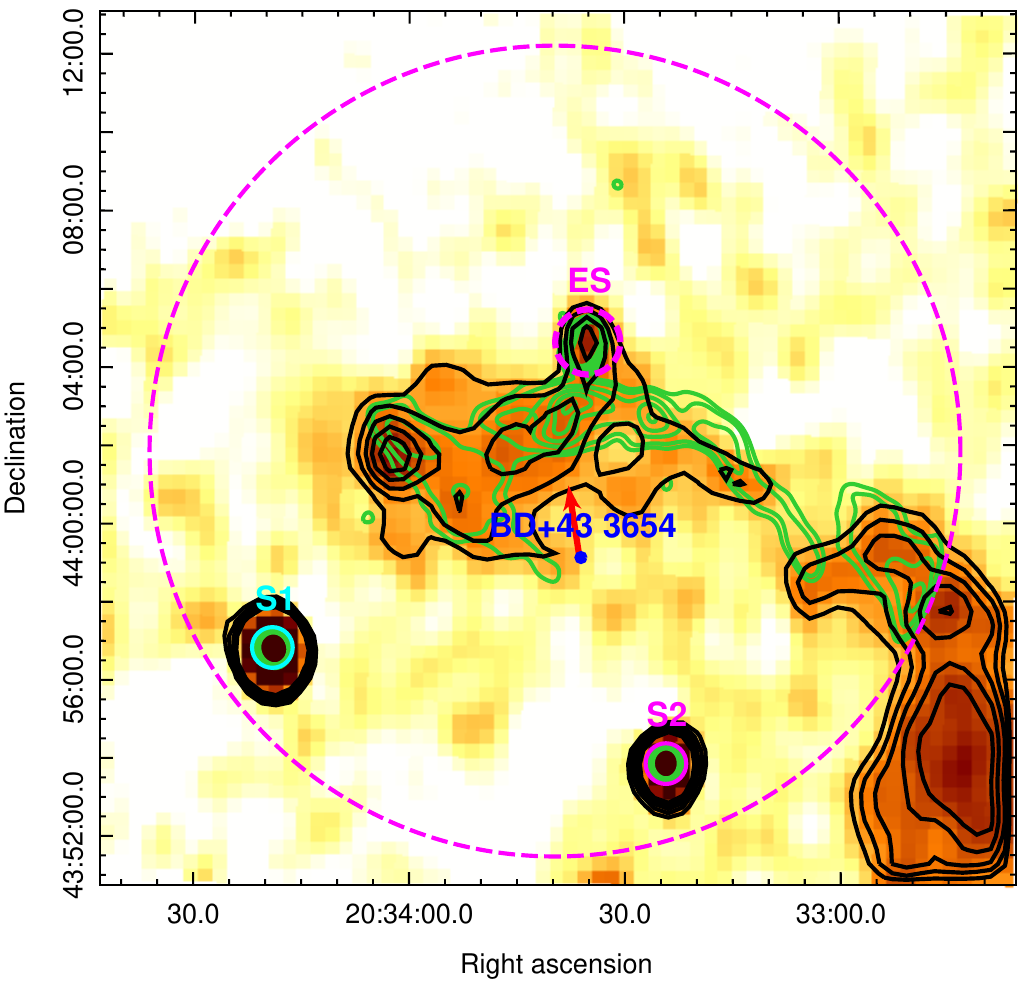}
    \caption{WENSS intensity image of \BD at 325~MHz. The full width at half maximum of the beam is $\approx 54''\times 77''$. The black contour levels are at 44, 54, 64, 74, and 84~\mJybeam. We overplot green contours from the intensity map at 3~GHz from \cite{Benaglia_2021} at the levels of 1.5, 2, 3, 4, 5, and 6~\mJybeam; the dashed circle corresponds to the primary beam of that observation.}
    \label{fig:WENSS_BD43}
\end{figure}

\begin{figure}
    \centering
    \includegraphics[width=\linewidth]{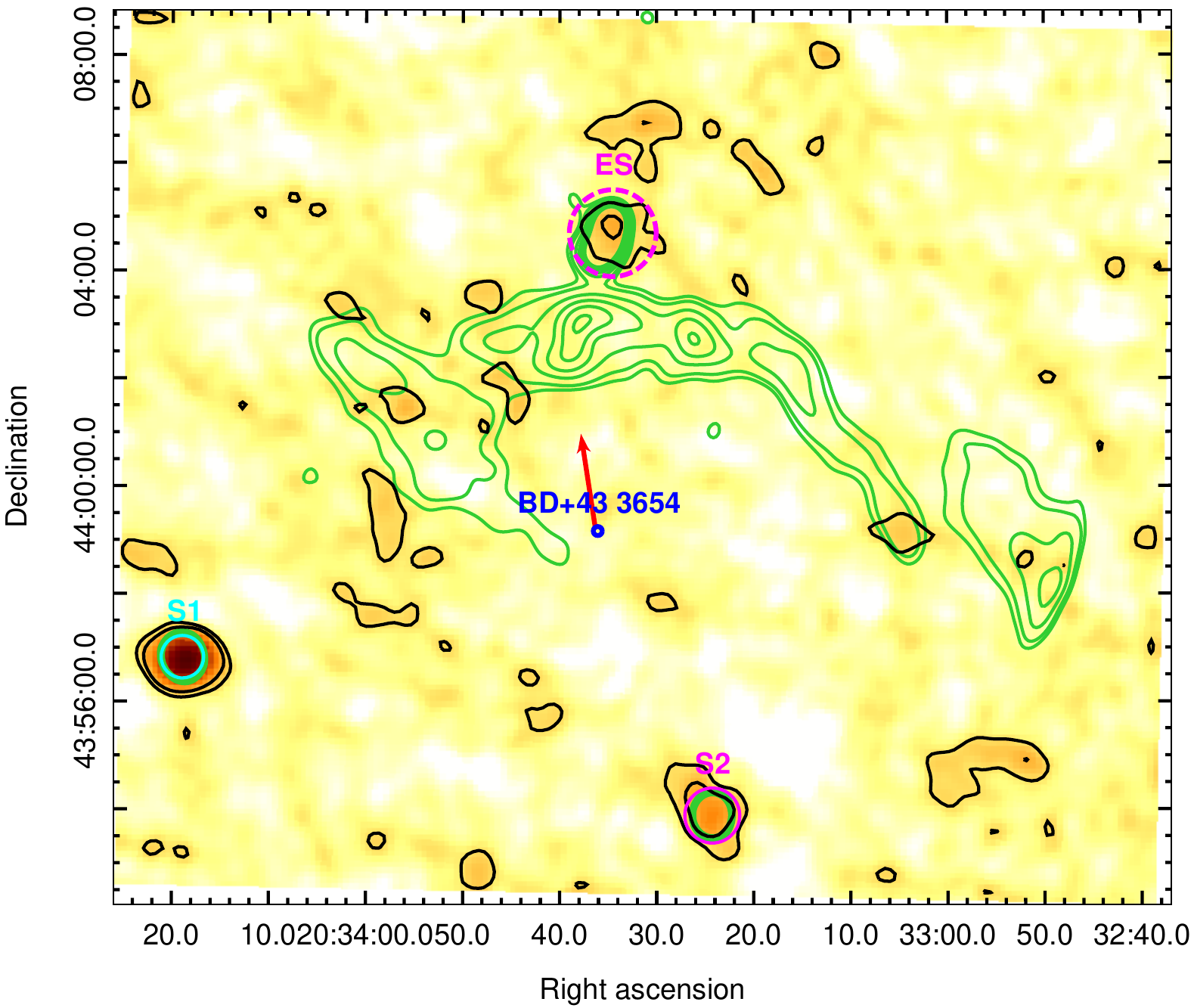}
    \caption{Intensity map of \BD at 150~MHz. The beam size is $25''\times 25''$. The black contour levels are at 10 and 20~\mJybeam. We overplot the same contours as in Fig.~\ref{fig:WENSS_BD43}.}
    \label{fig:TGSS_BD43}
\end{figure}

Here we present the radio maps at 150~MHz from the TGSS survey \citep{Intema_2017} and at 325~MHz from the WENSS survey \citep{WENSS_1997}, as they have not been discussed in detail in the literature. We note that the WENSS map for \BD was shown in \cite{Brookes_2016}, but the more recent radio data from \cite{Benaglia_2021} allow us to show a more insightful comparison (Fig.~\ref{fig:WENSS_BD43}). We used CARTA\footnote{\url{https://cartavis.org/}} to integrate the flux within the contour in which significant emission from the BS is seen. The total flux (excluding the ES source corresponding to an HII region) is $\approx 1.5$~Jy. We note that the map in Fig.~\ref{fig:WENSS_BD43} seems to suggest the presence of a point-like source to the left side of the BS; excluding it, the integrated flux is $\approx 1$~Jy. Given the low signal-to-noise in this map and the poor angular resolution, we refrained from over-interpreting these fluxes, but simply state that the total flux from \BD at 325~MHz is within 1--1.5~Jy. For completeness, we also show the map at 150~MHz, though no significant emission is detected as the map is not very deep ($rms \approx 7$~\mJybeam; Fig.~\ref{fig:TGSS_BD43}). 

\begin{figure}
    \centering
    \includegraphics[width=\linewidth]{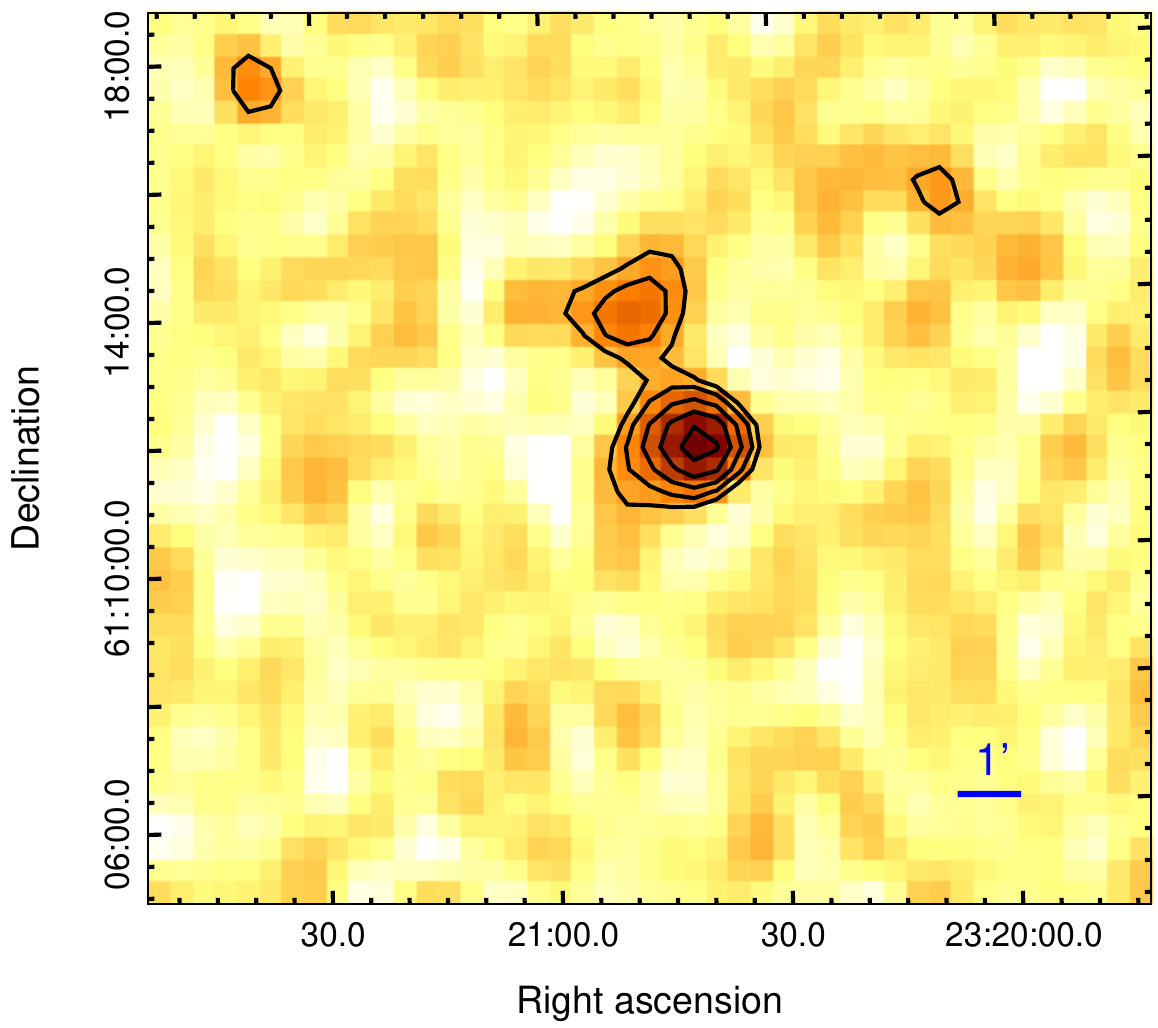}
    \caption{WENSS intensity image of \BN. The brightest source (south) corresponds to the unresolved emission from the BS of \BN together with the dense ionised clumps (Sect.~\ref{subsec:BD60}); the northern arc-shaped source is not from the BS itself, and is also detected in the 4--12~GHz map by \cite{Moutzouri_2022}. The map has $rms = 10$~\mJybeam and an angular resolution of $54''\times62''$. Contour levels are at 30, 50, 80, 120, and 170~\mJybeam.}
    \label{fig:WENSS_BD60}
\end{figure}

In the case of \BN, the WENSS map resembles that from \cite{Moutzouri_2022}, although with a poorer angular resolution. The dense ionised clumps close to the \BN are not resolved from the BS emission. The peak flux from this region is 217~\mJybeam (Fig.~\ref{fig:WENSS_BD60}).

\section{H$_\alpha$ emission of Vela X-1} \label{appendix:H_alpha}

\begin{figure}
    \centering
    \includegraphics[angle=270, width=0.49\textwidth]{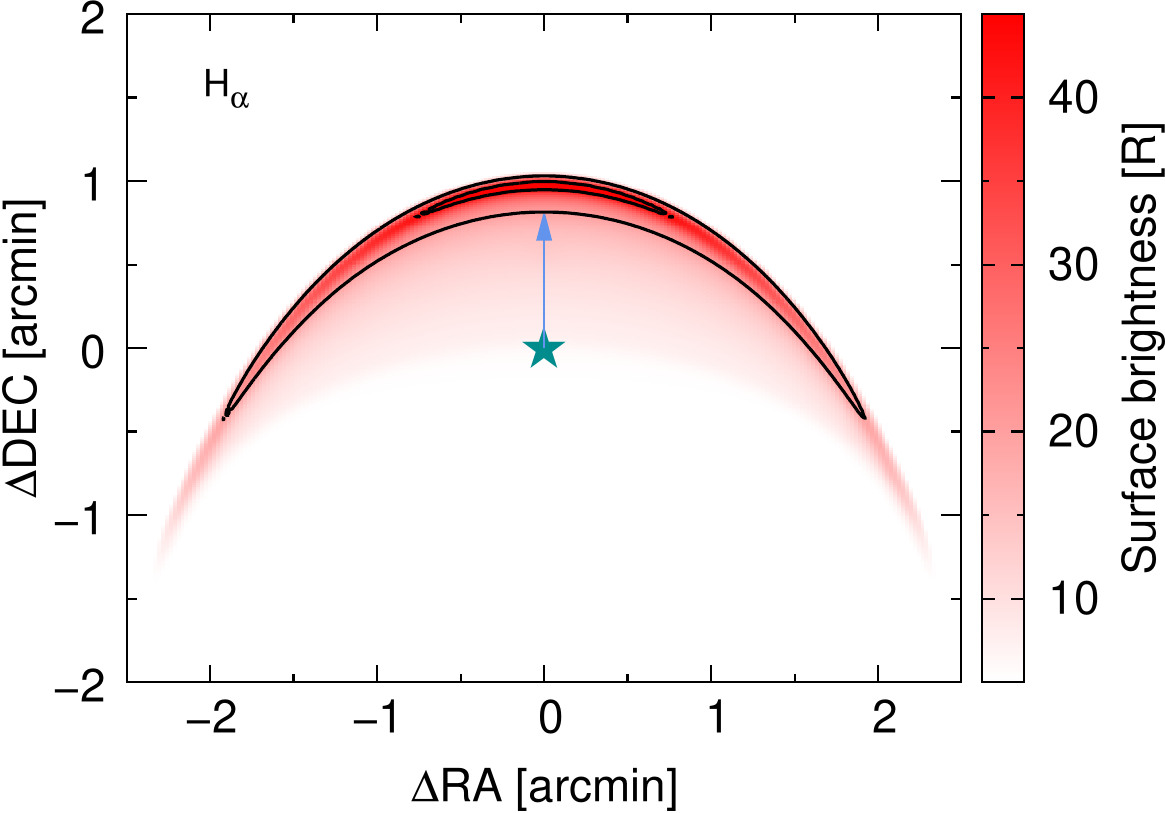}
    \caption{Simulated H$_\alpha$ surface brightness with our model. Contour levels are 20 and 43~R.}
    \label{fig:Vela_H_alpha}
\end{figure}

The ionised gas in the IL is also responsible for producing the H$_\alpha$ emission from the stellar BS. Thus, this poses an additional constraint to our model that is sensitive only to the width of the IL via the free parameter $\eta_H$. We therefore computed the H$_\alpha$ surface brightness map of Vela X-1 and compared it with the observed one from \cite{Gvaramadze_2018} to estimate the width of the IL. We calculated the H$_\alpha$ emission adapting the expression given by \cite{Gvaramadze_2018} as
\begin{equation}
    S_\alpha(\theta) \approx 4.3 \times 10^{10} \frac{\kappa(\theta)\,{\rm d}V(\theta)}{d^2}\; {\rm erg\,s^{-1}\,cm^{-2}},
\end{equation}
with d$V(\theta) \propto H_{\rm IL}(\theta)$ the volume of each cell, and
\begin{equation}
    \kappa(\theta) = 2.85 \times 10^{33}\,\left(\frac{T}{{\rm K}}\right)^{-0.9}\left(\frac{n_e}{{\rm cm}^{-3}}\right)\,\left(\frac{n_p}{{\rm cm}^{-3}}\right). 
\end{equation}
We finally convolved the synthetic map with a circular beam and divided by the beam area to get a surface brightness map in units of $\mathrm{R} = 5.66 \times 10^{-18}~$erg\,s$^{-1}$\,cm$^{-2}$\,arcsec$^{-2}$.  
\cite{Gvaramadze_2018} measured a peak surface brightness of 43~R near the apex of the BS of Vela X-1. To match this value, we needed to set $\eta_{\rm H} \approx 0.1$ (Fig.~\ref{fig:Vela_H_alpha}), which implies a significant reduction in the IL. 

\end{document}